\documentclass[twocolumn,amssymb, nobibnotes, aps, prb]{revtex4-2}

\usepackage{url}
\usepackage{bm}
\usepackage{graphicx}
\usepackage{amsmath}
\usepackage{amstext}
\usepackage{amsbsy}
\usepackage{color}
\usepackage[colorlinks=true, urlcolor=blue, linkcolor=blue, citecolor=blue, pdftex]{hyperref}
\usepackage{multirow}
\usepackage{float}
\usepackage[utf8]{inputenc}

\usepackage{tikz}

\newcommand{\plaquette}{%
	\mathord{%
		\begin{tikzpicture}[baseline=0.ex, line width=0.3pt, scale=0.7]
			\draw (0,0) rectangle (0.18,0.18);
			\draw (0,0) -- (0.18,0.18);
			\draw (0,0.18) -- (0.18,0);
		\end{tikzpicture}%
	}%
}

\definecolor{karlocolor}{rgb}{0.9,0.2,0}
\definecolor{pepecolor}{rgb}{0.9,0.05,0.05}
\definecolor{wipcolor}{rgb}{0.5,0.0,0.5}
\definecolor{ChatGPTcolor}{rgb}{0.0,0.5,0.5}

\newcommand{\ket}[1]{{\left| #1 \right\rangle}}
\newcommand{\bra}[1]{{\left \langle #1 \right|}}
\begin{document}

\title{Rigorous Anderson-type lower bounds on the ground-state energy of the pyrochlore Heisenberg antiferromagnet
}

\author{Péter Kránitz}
\affiliation{Institute of Physics,
Budapest University of Technology and Economics, M\"uegyetem rkp. 3., H-1111 Budapest, Hungary}
\affiliation{Institute for Solid State Physics and Optics, HUN-REN Wigner Research Centre for Physics, H-1525 Budapest, P.O.B.~49, Hungary}

\author{Karlo Penc}
\affiliation{Institute for Solid State Physics and Optics, HUN-REN Wigner Research Centre for Physics, H-1525 Budapest, P.O.B.~49, Hungary}

\date{\today}

\begin{abstract}

We construct rigorous Anderson-type lower bounds on the ground-state energy of the spin-$S$ Heisenberg antiferromagnet on the pyrochlore lattice.
By formulating and optimizing a hierarchy of local cluster motifs ordered by size, we generate a sequence of increasingly tight bounds.
A seven-site “hourglass” cluster composed of two corner-sharing tetrahedra furnishes an optimal lower bound that admits a closed-form expression for arbitrary spin $S$.
We also derive exact lower bounds for generalized models with further-neighbor exchange, ring exchange, and scalar spin-chirality interactions.
For $S=1/2$ and $S=1$, numerical optimization of an 18-site “crown” cluster containing a hexagonal loop yields rigorous lower bounds on the ground-state energy per site of the nearest-neighbor Heisenberg model with unit exchange, $e_\mathrm{GS} \geq -0.549832$ and $ \geq -1.632985$, respectively.
We compare the resulting bounds with numerical ground-state energy estimates from the literature.

\end{abstract}

\maketitle

\section{Introduction}

Determining the ground-state properties of quantum antiferromagnets remains a central challenge in condensed-matter physics.
In contrast to ferromagnets, where the fully polarized state is an exact eigenstate of the Heisenberg Hamiltonian, classical  Néel--ordered states in antiferromagnets are generally not eigenstates of the corresponding quantum problem.
Nevertheless, for many unfrustrated bipartite lattices, the Néel state provides a helpful starting point for understanding the ground state, raising the question of how such classical intuition can be justified in a controlled and quantitative manner.

In his 1951 work \cite{Anderson}, Anderson devised a construction that provides a rigorous lower bound on the ground-state energy by decomposing the lattice Hamiltonian into a sum of local sub-Hamiltonians and applying the variational principle to each term.
At the same time, simple product states provide natural variational upper bounds.
Together, these bounds constrain the exact ground-state energy and provide a quantitative justification for semiclassical approximations in bipartite quantum antiferromagnets.

The essence of Anderson’s construction is simple and general.
We write the Hamiltonian as a sum of sub-Hamiltonians
\begin{equation}
  \mathcal{H} = \sum_{i=1}^{M} \mathcal{H}_i ,
  \label{eq:lowb}
\end{equation}
where $M$ denotes the number of sub-Hamiltonians and their support includes a finite number of sites.
Denoting by $\ket{\Psi_\mathrm{GS}}$ the exact (and not necessarily known) ground state of $\mathcal{H}$ with energy $E_\mathrm{GS}$, we may use $\ket{\Psi_\mathrm{GS}}$ as a trial state for each sub-Hamiltonian.
The variational principle then implies
\begin{equation}
  E_0^{(i)} \leq \bra{\Psi_\mathrm{GS}} \mathcal{H}_i \ket{\Psi_\mathrm{GS}} ,
\end{equation}
where $E_0^{(i)}$ denotes the ground-state energy of $\mathcal{H}_i$.
Summing over all sub-Hamiltonians and using Eq.~(\ref{eq:lowb}), we arrive at
\begin{equation}
  \sum_{i=1}^{M} E_0^{(i)} \le E_\mathrm{GS},
  \label{eq:lower_bound_sum}
\end{equation}
which provides a rigorous lower bound on the exact ground-state energy.
The usefulness of this bound depends crucially on the choice of the local clusters and on how faithfully they capture the relevant correlations of the underlying lattice.

On frustrated lattices, such as the pyrochlore lattice considered here, classical long-range order is absent \cite{Moessner_PhysRevLett.80.2929}.
In this setting, lower-bound constructions are particularly valuable because they provide rigorous energy constraints without assuming any ordering pattern.
Refining Anderson’s idea for strongly frustrated systems, therefore, offers a systematic route to benchmark approximate analytical and numerical approaches and to test the compatibility of proposed local correlations with the global ground state.

Anderson-type lower bounds have a long history in quantum magnetism and have been refined in a variety of settings.
Early applications include the one-dimensional $J_1$--$J_2$ Heisenberg chain \cite{Majumdar69} and the triangular-lattice Heisenberg antiferromagnet \cite{Majumdar76}.
Subsequent work improved square-lattice bounds by employing more general cluster decompositions \cite{Tarrach90} and by optimizing modified exchange parameters within the sub-Hamiltonians \cite{Nishimori89}.
More recently, relaxations of the ground-state problem—such as semidefinite and bootstrap-type methods—have produced tight certificates with favorable scaling \cite{Barthel2012,han2020,Requena2023,Wang2024,Kull2024}.
In parallel, motif-based Anderson constructions remain valuable as a geometrically transparent framework that can be systematically improved by enlarging and optimizing the chosen clusters.

\begin{figure}[tb]
	\centering
	\includegraphics[width=0.6\linewidth]{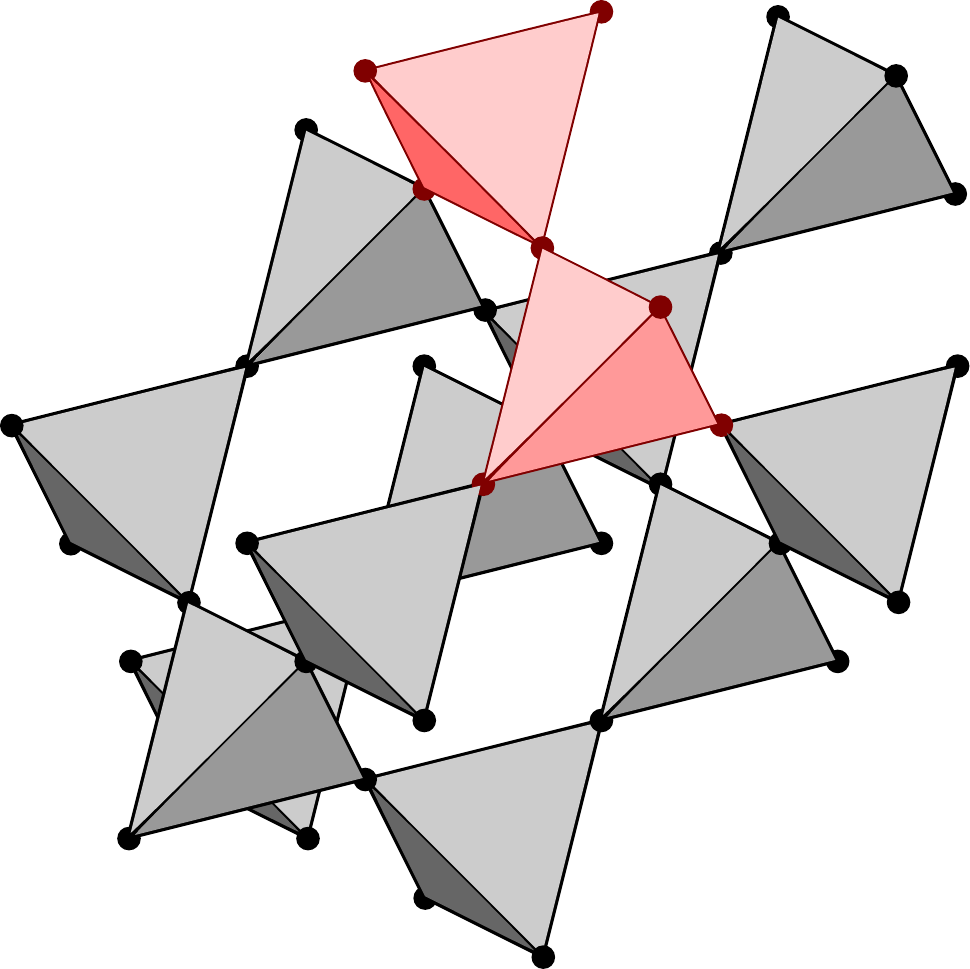}
	\caption{Pyrochlore lattice consisting of corner-sharing tetrahedra, with an embedded hourglass cluster (highlighted in red) serving as a representative motif for defining the sub-Hamiltonian $\mathcal{H}_i$. In this case, there is an hourglass for each site ($M=N$ in Eq.~(\ref{eq:lower_bound_sum})).}
	\label{fig:pyrochlore_lattice}
\end{figure}

In this work, we apply and extend Anderson’s lower-bound approach to the nearest-neighbor spin-$S$ Heisenberg antiferromagnet on the pyrochlore lattice,
\begin{equation}
  \mathcal{H} = J \sum_{\langle i,j \rangle} \mathbf{S}_i \cdot \mathbf{S}_j ,
  \label{eq:heisenberg_model}
\end{equation}
where $J>0$ and $\langle i,j\rangle$ denotes nearest-neighbor pairs.
The pyrochlore lattice is a three-dimensional network of corner-sharing tetrahedra and represents a paradigmatic example of strong geometric frustration. For $S=\tfrac12$ in particular, the nature of the ground state remains under active debate, 
with proposals ranging from quantum spin-liquid phases \cite{canals,yasir,pohle2023} 
to various valence-bond and symmetry-broken states \cite{hering2022,hagymasi,astrakhantsev2021,schafer,cheng2025}.
High-precision numerical and semi-analytical studies on frustrated quantum magnets, including coupled-cluster analyses pioneered by Richter and collaborators, 
provide important reference points for the ground-state energy in such systems \cite{richter2004,2004JPCM...16S.779R}.

The remainder of the paper is organized as follows.
In Sec.~\ref{sec:simple_bounds}, we introduce Anderson-type lower bounds for the pyrochlore lattice using analytically tractable sub-Hamiltonians, including Anderson’s original site-centered star construction and single-tetrahedron clusters.
In Sec.~\ref{sec:hourglass}, we introduce the seven-site hourglass motif formed by two corner-sharing tetrahedra and show how optimizing exchange parameters within this cluster yields systematically improved lower bounds, including a closed-form expression valid for arbitrary spin $S$.
Extensions of the hourglass construction to more general Hamiltonians, including further-neighbor exchange interactions, ring exchange, and scalar spin-chirality terms, are discussed in Sec.~\ref{sec:generalizations}.
In Sec.~\ref{sec:crown}, we move beyond the seven-site cluster and study the 18-site crown motif, which explicitly incorporates hexagonal loops. We present brute-force results for spin-$\tfrac12$ Heisenberg and chiral models and introduce a controlled Hilbert-space reduction scheme that enables optimized bounds for the $S=1$ Heisenberg model.
Finally, Sec.~\ref{sec:conclusions} summarizes our findings and outlines possible directions for further extensions of the approach.
Technical details, explicit spectra, and the complete form of the crown-cluster Hamiltonian are provided in the Appendices.

\section{Simple bounds}
\label{sec:simple_bounds}

\subsection{Anderson's original star construction}
\label{sec:star}

\begin{table*}[bt]
\centering
\caption{
\label{tab:bounds_comparison_transposed}
Comparison of Anderson-type lower bounds on the ground-state energy of the nearest-neighbor Heisenberg antiferromagnet on the pyrochlore lattice.
Entries show the lower bounds $e_\mathrm{LB}/J$ obtained from different choices of sub-Hamiltonians: the site-centered Anderson star (Sec.~\ref{sec:star}), the single-tetrahedron cluster (Sec.~\ref{sec:tetra}), the seven-site hourglass motif (Sec.~\ref{sec:analytic_hourglass}), and two variants of the 18-site crown cluster (Sec.~\ref{sec:crown}).
The entry labeled ``Crown*'' corresponds to Eq.~(\ref{eq:crown_red}), the reduced crown construction with additional constraints enforcing conserved outer-pair spin quantum numbers, while ``Crown'' denotes the unconstrained crown optimization given the Hamiltonian (\ref{eq:crown_hamiltonian}).
Exact analytic expressions are listed where available.
}
\begin{ruledtabular}
\begin{tabular}{lccccc}
 & Star & Tetrahedron & Hourglass (7-site) & Crown* (18 sites) & Crown (18 sites) \\
\hline
Exact expression 
& $-\dfrac{1}{2}S(6S+1)$
& $-S(S+1)$
& $-S(S+1) \dfrac{4S+1}{4S+2}$ 
\\[6pt]
$S=\frac{1}{2}$ 
& $-1 $
& $-\frac{3}{4} = -0.75$
& $-\frac{9}{16} = -0.5625$ 
& $-0.550158(2)$
& $-0.549832(8)$
\\[6pt]
$S=1$ 
& $-\frac{7}{2} = -3.5$
& $-2$ 
& $-\frac{5}{3} \approx -1.6667$ 
& $-1.632985(8)$ &
\\[6pt]
$S=\frac{3}{2}$ 
& $-\frac{15}{2} = -7.5$
& $-\frac{15}{4} = -3.75$
& $-\frac{105}{32} = -3.2815$ 
\\[6pt]
$S\to\infty$ 
& $-3S^2 - \frac{1}{2} S$ 
& $-S^2 - S$ 
& $-S^2 - \tfrac34 S + O(1)$ 
\end{tabular}
\end{ruledtabular}
\end{table*}

\begin{figure}[tb]
	\centering
	\includegraphics[width=0.9\linewidth]{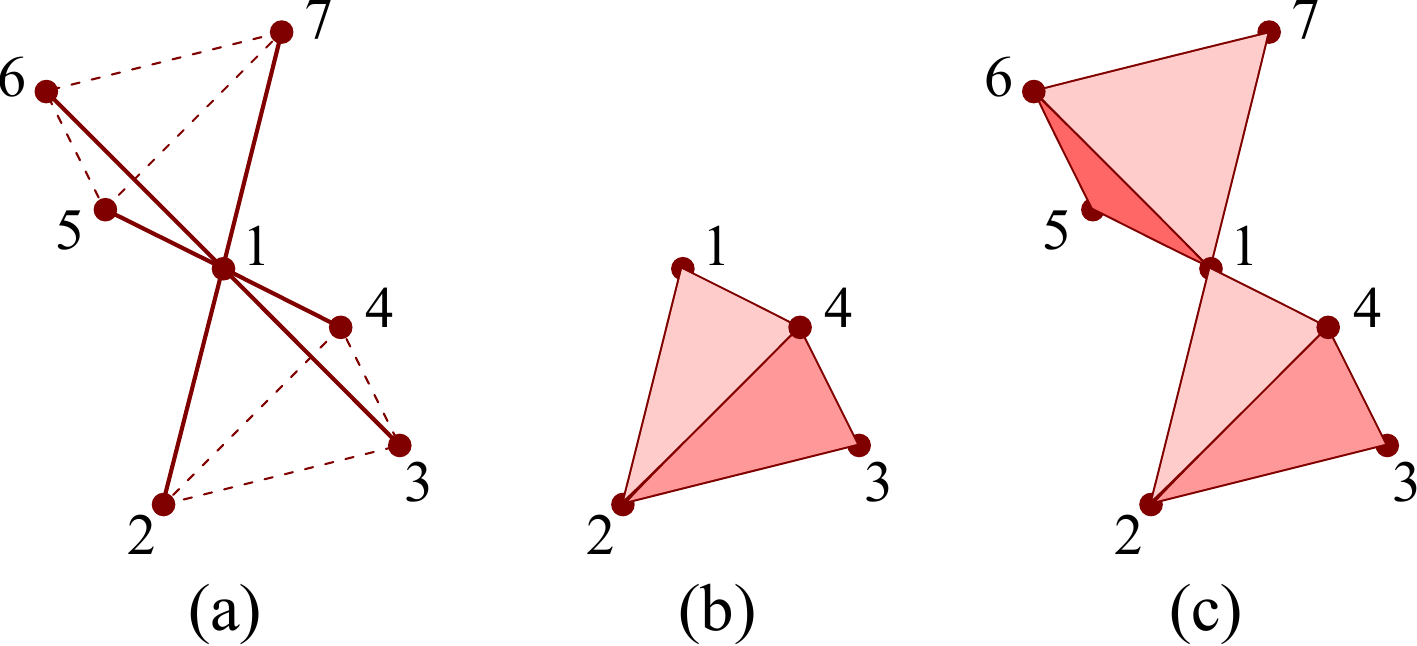}
	\caption{Clusters used to derive rigorous lower bounds on the ground‑state energy.  
(a) The star cluster of Anderson’s original argument, comprising a central spin (labelled `1') coupled to its neighbouring spins (sites `2'--`7').  
(b) The tetrahedron cluster, which incorporates the intrinsic frustration of the pyrochlore lattice and therefore gives a tighter bound.  
(c) The seven‑site hourglass cluster, which merges the star and tetrahedron motifs and further improves the lower bound.}
	\label{fig:hourglass_star_tatra}
\end{figure}

Following Anderson's original star construction, we decompose the Hamiltonian (\ref{eq:heisenberg_model}) into a sum of site-centered star Hamiltonians, shown in Fig.~\ref{fig:hourglass_star_tatra}(a). 
Since summing stars over all sites double-counts each bond, we assign half of each bond to each endpoint, 
\begin{equation}
	\mathcal{H}_i = \frac{J}{2}\sum_{j\in \mathrm{NN}(i)} \mathbf{S}_i \cdot \mathbf{S}_j ,
	\label{eq:star_decomp}
\end{equation}
where $\mathrm{NN}(i)$ denotes the $Z$ nearest neighbors of site $i$ (for the pyrochlore lattice, $Z=6$). There are as many star sub-Hamiltonians as sites in a periodic lattice. The energy of the star Hamiltonian $\mathcal{H}_i$ is minimized by maximizing the total spin of the neighbor shell to $ZS$ and coupling it antiparallel to the central spin, yielding
\begin{equation}
	E_0^{(i)} = -\frac{J}{2}\,S\,(ZS+1).
\end{equation}
 Summing over all sites, the Anderson-star lower bound for the pyrochlore lattice becomes 
\begin{equation}
	\frac{E_\mathrm{GS}}{N}  \geq -\frac{J}{2}\,S\,(6S+1).
\end{equation}
In particular, one finds $E_\mathrm{GS}/N \ge -J$ for $S=\tfrac{1}{2}$ and  $E_\mathrm{GS}/N \ge -7J/2 = -3.5 J$ for $S=1$. We collected these values in Tab.~\ref{tab:bounds_comparison_transposed}. 
We note that the star construction neglects geometric frustration, resulting in a lower bound that scales as $\propto -3S^2$ in the semiclassical limit and is therefore far from the exact ground-state energy.

\subsection{Tetrahedral cluster}
\label{sec:tetra}

A more natural choice of sub-Hamiltonians for the pyrochlore lattice is provided by individual tetrahedra, which form the fundamental building blocks of the lattice.
For a tetrahedron $(i)$ with sites $(i_1,i_2,i_3,i_4)$, the corresponding sub-Hamiltonian may be written as
\begin{equation}
	\mathcal{H}_i=\frac{J}{2}\left(\mathbf{S}_{i_1}+\mathbf{S}_{i_2}+\mathbf{S}_{i_3}+\mathbf{S}_{i_4}\right)^2-2S(S+1)J
\end{equation} 
whose ground-state energy is $E_0^{(i)}=-2J S(S+1)$, corresponding to a total-spin singlet.
 Since there are $M=N/2$ tetrahedra in a periodic $N$-site pyrochlore lattice, this yields the lower bound
\begin{equation}
  e_\mathrm{LB} = -S(S+1)J \;.
\end{equation}
For spin-1/2, this gives $E_\mathrm{GS}/N \ge -0.75 J$, and for spin-1 $E_\mathrm{GS}/N \ge -2 J$. 
These bounds are tighter than those from the star construction, in the sense that they lie closer to the exact ground-state energy from below. Also, the tetrahedral construction reproduces the correct $\propto -S^2$ semiclassical energy.

\section{7-site motif: hourglass}
\label{sec:hourglass}

\begin{figure}[bt]
	\centering
	\includegraphics[width=0.9\linewidth]{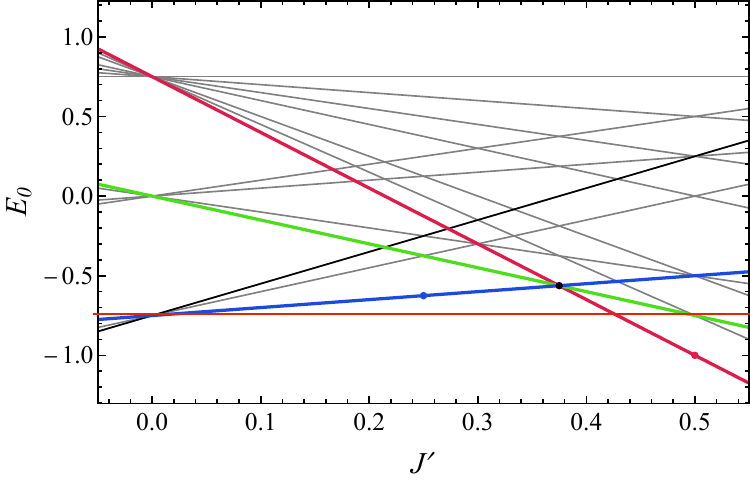}
	\caption{The energy spectrum of the 7-site hourglass subHamiltonian, Eq.~(\ref{eq:motifham7_nn}), as a function of the free parameter $J'$ (we set $J=1$). The optimal lower bound $-0.5625$ is achieved for $J'=J^*=3/8$, where the energies of the different irreducible representations cross. For $J'=0.5$, we recover the lower bound (red circle) from the star-cluster (Sec.~\ref{sec:star}), while for $J' = 1/4$, the blue circle indicates the case of equal exchange couplings on all bonds.}
	\label{fig:7_site_motif}
\end{figure}

The constructions discussed above highlight complementary aspects of Anderson’s lower-bound approach.
The star cluster captures the local competition between a single spin and its environment. Still, it largely neglects frustration among neighboring spins, resulting in a relatively weak bound on the pyrochlore lattice.
By contrast, the tetrahedral cluster incorporates the lattice's fundamental frustrated unit and imposes local singlet constraints, leading to a substantially improved bound.
These observations suggest that further improvement requires a cluster that simultaneously captures both local frustration and the coupling of a tetrahedral unit to its surroundings.

We therefore introduce a seven-site ``hourglass'' motif, shown in Fig.~\ref{fig:hourglass_star_tatra}(c), formed by joining two tetrahedra at a common site.
This motif naturally combines the key features of the star and tetrahedral constructions.
In the following, we construct the corresponding sub-Hamiltonian and determine the resulting lower bound.

\subsection{The Sub-Hamiltonian Ansatz}

\begin{table}[tb]
\centering
\caption{\label{tab:D3d_char}
Character table for the $D_{3d}$ point group of the hourglass cluster. The $i$ is the inversion, represented as $(2\ 7)(3\ 6)(4\ 5)$ using cyclic permutations, $C_3$ is the three-fold rotation $(2\ 3\ 4)(5\ 7\ 6)$, $3C_2'$ is the two-fold rotation $(2\ 7)(3\ 5)(4\ 6)$, $\sigma_d$ is the reflection $(2\ 4)(5\ 7)$, and $S_6$ is the rotoreflection $(2\ 6\ 4\ 7\ 3\ 5)$.
}
\begin{ruledtabular}
\begin{tabular}{ccccccc}
$D_{3d}$ & $E$ & $2C_3$ & $3C_2'$ & $i$ & $2S_6$ & $3\sigma_d$ \\
\hline
$A_{1g}$ & 1 & 1 & 1 & 1 & 1 & 1  \\
$A_{2g}$ & 1 & 1 & -1 & 1 & 1 & -1 \\
$E_g$ & 2 & -1 & 0 & 2 & -1 & 0  \\ 
$A_{1u}$ & 1 & 1 & 1 & -1 & -1 & -1  \\
$A_{2u}$ & 1 & 1 & -1 & -1 & -1 & 1 \\
$E_u$ & 2 & -1 & 0 & -2 & 1 & 0 
\end{tabular}
\end{ruledtabular}
\end{table}

Anderson’s construction allows us to introduce additional exchange terms and interactions in the sub-Hamiltonians that do not appear explicitly in the lattice Hamiltonian, as long as these terms vanish when summed over all motifs.
This controlled freedom enables us to design sub-Hamiltonians that better capture local correlations.
We implement this strategy by exploiting the symmetries of the pyrochlore lattice and of the hourglass motif.

The Heisenberg Hamiltonian in Eq.~(\ref{eq:heisenberg_model}) respects the full space-group symmetry  ($S$) of the pyrochlore lattice, $Fd\bar{3}m$ (No.~227)~\cite{ITC_A}, with face-centered cubic (FCC) translation group  and octahedral point group $O_h$, which can be viewed as the extension of the tetrahedral group $T_d$ by inversion $C_i$. The space group is non-symmorphic due to fractional translations accompanying certain point-group operations.
Space-group invariance implies that sub-Hamiltonians associated with symmetry-related motifs are unitarily equivalent.
For any two motifs $i$ and $j$ related by a space-group operation $g \in S$, we have
\begin{equation}
\mathcal{H}_i = g \mathcal{H}_j g^{-1}.
\label{eq:hamiltonian_restriction}
\end{equation}
When this condition holds, all sub-Hamiltonians are identical up to symmetry, and it suffices to consider a single representative sub-Hamiltonian $\mathcal{H}_0$ with ground-state energy $E_0$.
Equation~(\ref{eq:lower_bound_sum}) then reduces to
\begin{equation}
  \label{eq:eLBMN}
 e_{\mathrm{LB}}\equiv \frac{M}{N} E_0 \le  \frac{E_{\mathrm{GS}}}{N},
\end{equation}
where $N$ is the total number of lattice sites and $M$ is the number of motifs covering the lattice. Furthermore, each local sub-Hamiltonian $\mathcal{H}_i$ inherits the point-group symmetry of the corresponding cluster and is therefore invariant under a subgroup of the cubic point group $O_h$.

In particular, the seven-site motif is invariant under the point group $D_{3d}$, which has order 12.
As a consequence, the corresponding sub-Hamiltonian $\mathcal{H}_0$ commutes with all elements of $D_{3d}$,
\begin{equation}
[\mathcal{H}_0, g] = 0, \qquad \forall\, g \in D_{3d}.
\end{equation}

\subsection{Bilinear hourglass Hamiltonian}

In principle, an arbitrary seven-site Hamiltonian may contain multi-spin interactions and higher-order terms, especially for $S\geq 1$.
Here we restrict ourselves to bilinear, SU(2)-invariant interactions of the form $\mathbf{S}_i\cdot\mathbf{S}_j$. This choice is minimal but directly connects to the underlying Heisenberg model and is already sufficient to yield a nontrivial improvement over Anderson’s lower bound. 

The $D_{3d}$-symmetric sub-Hamiltonian for the seven-site motif then takes the form
\begin{multline}
	\mathcal{H}_\text{0}(J,J^\prime) = 
	J^\prime\left( \mathbf{S}_1\cdot\mathbf{S}_2 + \mathbf{S}_1\cdot\mathbf{S}_3+ \mathbf{S}_1\cdot\mathbf{S}_4 \right.
	\\
	+ \left. \mathbf{S}_1\cdot\mathbf{S}_5+ \mathbf{S}_1\cdot\mathbf{S}_6+ \mathbf{S}_1\cdot\mathbf{S}_7
	\right)
	\\
	+\left(\frac{J}{2}-J^\prime\right)\left(\mathbf{S}_2\cdot \mathbf{S}_3+\mathbf{S}_2\cdot \mathbf{S}_4+\mathbf{S}_3\cdot \mathbf{S}_4\right)
	\\
	+\left(\frac{J}{2}-J^\prime\right)\left( \mathbf{S}_5 \cdot\mathbf{S}_6+\mathbf{S}_5\cdot \mathbf{S}_7+\mathbf{S}_6 \cdot\mathbf{S}_7\right),
	\label{eq:motifham7_nn}
\end{multline}
where $J^\prime$ acts as a free parameter. 
We determine the optimal lower bound by maximizing the ground-state energy,
\begin{equation}
	e_{\mathrm{LB}} (J) =\max_{J^\prime} E_0(J,J^\prime),
\end{equation}
using the fact that $M=N$ for the hourglass embedding [Eq.~(\ref{eq:eLBMN})].
This hourglass motif comprises both previous constructions as special cases and therefore implies a lower bound that is at least as tight as those obtained from either of them. Choosing $J^\prime=J/2$ reproduces Anderson’s original star cluster (Sec.~\ref{sec:star}), while the tetrahedral construction is recovered upon relaxing the requirement of $D_{3d}$ point-group symmetry of the sub-Hamiltonian.

\subsection{Direct diagonalization for $S=1/2$}
\label{sec:direct_diag_hourglass}

\begin{table}[t]
  \caption{Irreducible-representation content of seven $S=\tfrac12$ spins under the $D_{3d}$ point group, resolved by total spin $S_{\mathrm{tot}}$. Table entries give the multiplicity $m_\Gamma$ of each irreducible representation in a given spin sector. The number of states contributed by a given entry is therefore $m_\Gamma \times \dim(\Gamma)$, and the total number of many-body states in each $S_{\mathrm{tot}}$ sector is obtained by multiplying further by the spin degeneracy $(2S_{\mathrm{tot}}+1)$.	
	\label{tab:D3d_7spins}}
	\begin{ruledtabular}
	\begin{tabular}{c c c c c c}
		Irrep $\Gamma$ & $\mathrm{dim}(\Gamma)$ 
		& $S_{\mathrm{tot}}=\tfrac12$
		& $S_{\mathrm{tot}}=\tfrac32$
		& $S_{\mathrm{tot}}=\tfrac52$
		& $S_{\mathrm{tot}}=\tfrac72$ \\
		\hline
		$A_{1g}$ & 1 & 2 & 2 & 1 & 1 \\
		$A_{2g}$ & 1 & 1 & 0 & 0 & 0 \\
		$E_g$    & 2 & 2 & 3 & 1 & 0 \\
		$A_{1u}$ & 1 & 1 & 1 & 0 & 0 \\
		$A_{2u}$ & 1 & 2 & 1 & 1 & 0 \\
		$E_u$    & 2 & 2 & 2 & 1 & 0 \\
		\hline
		Total multiplets &  & 14 & 14 & 6 & 1 \\
	\end{tabular}
	\end{ruledtabular}
\end{table}

For spin-$\tfrac12$, we determine the optimal lower bound by direct diagonalization of the hourglass Hamiltonian.
This procedure is entirely general and essentially algorithmic:
Once the sub-Hamiltonian is fixed, symmetry alone determines the block structure, and the spectrum follows from straightforward diagonalization.
No additional physical insight or model-specific assumptions are required.

We exploit global SU(2) spin-rotation symmetry together with the  $D_{3d}$ point-group symmetry of the cluster to reduce the Hamiltonian to independent blocks that can be diagonalized exactly.
In Table~\ref{tab:D3d_7spins}, we list the decomposition of the Hilbert space of seven $S=1/2$ spins into irreducible representations of $D_{3d}$, resolved by total spin $S_{\mathrm{tot}}$.
The highest multiplicity of any spatial irrep is three, implying that the largest symmetry block has dimension at most $3\times 3$ once the spin degeneracy is factored out.

 The symmetry-based direct diagonalization procedure outlined above provided us with the energy levels shown in Figure~\ref{fig:7_site_motif}. We find that only three total-spin sectors contribute to the ground state: $S_{\mathrm{tot}}=\tfrac12$, $\tfrac32$, and $\tfrac52$.
The corresponding levels are distinguished by color in the figure.

The blue branch corresponds to the $S_{\mathrm{tot}}=\tfrac12$ sector and includes states transforming according to the $A_{1g}$, $A_{1u}$, and $E_g$ irreducible representations.
Its energy varies linearly as
\begin{subequations}
\begin{equation}
E_0 = -\frac{3}{4}J + \frac{1}{2}J^\prime .
\end{equation}
The green branch represents the $S_{\mathrm{tot}}=\tfrac32$ sector, associated with the $E_g$ and $E_u$ representations, and has energy
\begin{equation}
E_0 = -\frac{3}{2}J^\prime .
\end{equation}
Finally, the red branch corresponds to the fully symmetric $A_{1g}$ multiplet in the $S_{\mathrm{tot}}=\tfrac52$ sector, with energy
\begin{equation}
E_0 = \frac{3}{4}J - \frac{7}{2}J^\prime .
\end{equation}
\end{subequations}

As $J^\prime$ is varied, these three energy levels intersect at a single point,
$J^\prime = \frac{3}{8}$, 
where all three become degenerate.
At this crossing, the common energy furnishes the optimal lower bound
\begin{equation}
e_\mathrm{LB} = -\frac{9}{16}J = -0.5625J.
\label{eq:hourglassLB_1/2}
\end{equation}
This value improves significantly upon the bounds obtained from the star cluster and the single-tetrahedron construction (see Table~\ref{tab:bounds_comparison_transposed} for detailed comparison).

All energy levels encountered in the spin-$\tfrac12$ hourglass arise from $1\times1$ symmetry blocks when performing the symmetry-based diagonalization outlined above.
A possible explanation is the presence of an enhanced symmetry.
For the nearest-neighbor Hamiltonian~(\ref{eq:motifham7_nn}), the symmetry of the problem is in fact higher than that of the geometric point group $D_{3d}$.
In particular, the sites of the upper and lower basal triangles can be permuted independently in the absence of longer ranged exchanges, and the two triangles can be exchanged.
Together, these operations generate a symmetry group isomorphic to 
\begin{equation}
 (\mathfrak{S}_3 \times \mathfrak{S}_3)\rtimes \mathbb{Z}_2\,,
  \label{eq:S3S3Z2}
\end{equation}
a subgroup of $\mathfrak{S}_6$ of order $72$ (here $\mathfrak{S}_3$ and $\mathfrak{S}_6$ denote the symmetric group of order $3!=6$ and $6!=720$, respectively).
However, it is not \emph{a priori} obvious that this symmetry alone is sufficient to enforce the complete absence of higher-dimensional symmetry blocks. 

We have applied the same symmetry-based diagonalization procedure to higher spin values.
Remarkably, the symmetry-resolved blocks remain one-dimensional for all spins considered.
This behavior is highly nontrivial: while the total Hilbert space grows rapidly with $S$, one would generally expect increasing multiplicities of irreducible representations.
The absence of such growth points to a deeper structural simplification of the hourglass spectrum and motivates the analytic formulation presented in the following subsection.

\subsection{Energy levels for arbitrary spins}
\label{sec:analytic_hourglass}

The hourglass Hamiltonian admits a formulation valid for arbitrary spin $S$.
We rewrite Eq.~(\ref{eq:motifham7_nn}) as
\begin{align}
	\mathcal{H}_0 =& \frac{J^\prime}{2}
	\left[\mathbf{S}_\mathrm{tot}^2 - \mathbf{S}_{234567}^2 - S(S+1)\right] \nonumber \\
	&+ \left(\frac{J}{4} - \frac{J^\prime}{2}\right)
	\left[\mathbf{S}_{234}^2 + \mathbf{S}_{567}^2 - 6S(S+1)\right],
	\label{eq:linearham}
\end{align}
where
\begin{subequations}
\label{eq:s234etc}
\begin{align}
\mathbf{S}_{234} &= \mathbf{S}_{2} + \mathbf{S}_{3} + \mathbf{S}_{4}, \\
\mathbf{S}_{567} &= \mathbf{S}_{5} + \mathbf{S}_{6} + \mathbf{S}_{7},
\end{align}
denote the total-spin operators of the two basal triangles,
\begin{equation}
\mathbf{S}_{234567} = \mathbf{S}_{234} + \mathbf{S}_{567},
\end{equation}
their combined total spin, and
\begin{equation}
\mathbf{S}_\mathrm{tot} = \mathbf{S}_{234567} + \mathbf{S}_{1},
\end{equation}
the total spin of the full seven-site cluster.
\end{subequations}
Written in terms of total-spin operators, $\mathcal H_0$ is manifestly invariant under the symmetry group $(\mathfrak{S}_3 \times \mathfrak{S}_3)\rtimes \mathbb{Z}_2$ defined in Eq.~(\ref{eq:S3S3Z2}).

Expressing the Hamiltonian solely in terms of the total-spin quantum numbers
$S_{234}$, $S_{567}$, $S_{234567}$, and $S_\mathrm{tot}$ reduces the problem to a finite set of analytically accessible energy levels. The resulting eigenvalues depend linearly on the coupling constants and require no explicit diagonalization of large matrices. This representation therefore provides a direct and computationally efficient route to rigorous lower bounds for arbitrary spin $S$.
  
We use the fact that the two triangles, $(234)$ and $(567)$, together with site 1 form a bipartite system.
The Lieb–Mattis theorem \cite{Lieb} then determines the spin multiplets of the ground state.
In particular, the total spin of the two triangles is maximal,
\begin{equation}
  S_{234567} = S_{234} + S_{567},
\end{equation}
while coupling to the central spin $S_1$ minimizes the total spin,
\begin{equation}
  S_\text{tot} = S_{234} + S_{567} - S.
\end{equation}
By investigating the possible total spin values on the (234) and (567) triangles, the lowest energies are:  
for $S_{234}=S_{567}=S$  
\begin{subequations}
\begin{equation}
 E_{0}^{(S,S)}= - J S(S+1) + S J^\prime \,,
\end{equation}
for $S_{234}=S, S_{567}=S+1$ and $S_{234}=S+1, S_{567}=S$ 
\begin{equation}
 E_{0}^{(S, S+1)} =  - \frac{1}{2} J (S+1) (2 S -1) - J' (S+1) \,,
\end{equation}
 and for $S_{234}=S+1, S_{567}=S+1$ 
\begin{equation}
E_{0}^{(S+1,S+1)} = - J (S-1) (S+1) - J' (3 S+2) \,.
\end{equation}
\end{subequations}
These energies cross at $J'=\frac{S+1}{4S+2}J$ (corresponding to the black dot at $J'=3/8$ in Fig.~\ref{fig:7_site_motif} for $S=1/2$), and the energy at the crossing point defines the lower bound
\begin{equation}
	e_\mathrm{LB}=-S(S+1)\frac{4S+1}{4S+2} .
\end{equation}
 We list some values in the Table~\ref {tab:bounds_comparison_transposed}.
Thus, considering the cluster of two connected tetrahedra, we can improve the lower bound based on a single tetrahedron by a factor of $\frac{4S+1}{4S+2}$.

\section{Generalizations of the Heisenberg model}
\label{sec:generalizations}

\subsection{$J_1$--$J_2$--$J_{3b}$ Heisenberg model}
\label{sec:J1J2J3}

We extend the nearest-neighbor Heisenberg model on the pyrochlore lattice by including further-neighbor interactions,
\begin{equation}
	\mathcal{H}
	=
	J_1 \sum_{\langle i,j \rangle} \mathbf{S}_i\cdot \mathbf{S}_j
	+J_2 \sum_{\langle\!\langle i,j \rangle\!\rangle} \mathbf{S}_i\cdot \mathbf{S}_j
	+J_{3b} \sum_{\langle\!\langle\!\langle i,j \rangle\!\rangle\!\rangle} \mathbf{S}_i\cdot \mathbf{S}_j ,
	\label{eq:pyroham_J1J2J3}
\end{equation}
where $J_1\equiv J$ denotes the nearest-neighbor exchange, $J_2$ the second-neighbor exchange, and $J_{3b}$ one of the two symmetry-inequivalent third-neighbor couplings on the pyrochlore lattice.
The sums $\langle i,j\rangle$ and $\langle\!\langle i,j\rangle\!\rangle$ run over nearest- and second-nearest-neighbor pairs, respectively, while
$\langle\!\langle\!\langle i,j\rangle\!\rangle\!\rangle$ denotes third-neighbor pairs belonging to adjacent corner-sharing tetrahedra and corresponds to the $J_{3b}$ exchange.
Among the two symmetry-inequivalent third-neighbor couplings, we keep only $J_{3b}$, which connects sites on adjacent corner-sharing tetrahedra, and exclude the alternative coupling across hexagons, commonly denoted $J_{3a}$.

Using the hourglass covering of the lattice, the presence of finite $J_2$ and $J_{3b}$ modifies the corresponding sub-Hamiltonian as
\begin{multline}
	\mathcal{H}_0(J_1,J_2,J^\prime) =
	\mathcal{H}_0(J_1,J^\prime)
\\
	+J_2\bigl(
	\mathbf{S}_2\cdot \mathbf{S}_5
	+\mathbf{S}_2\cdot \mathbf{S}_6
	+\mathbf{S}_3\cdot\mathbf{S}_5
	+\mathbf{S}_3\cdot\mathbf{S}_7
	+\mathbf{S}_4 \cdot\mathbf{S}_6
	+\mathbf{S}_4\cdot \mathbf{S}_7
	\bigr)
\\
	+J_{3b}\bigl(
	\mathbf{S}_2\cdot \mathbf{S}_7
	+\mathbf{S}_3\cdot\mathbf{S}_6
	+\mathbf{S}_4\cdot\mathbf{S}_5
	\bigr),
	\label{eq:motifham7_J1J2J3}
\end{multline}
where $\mathcal{H}_0(J_1,J^\prime)$ is the nearest-neighbor hourglass Hamiltonian defined in Eq.~(\ref{eq:motifham7_nn}).
Unlike nearest-neighbor bonds, each second- and third-neighbor bond appears only once in the hourglass covering.
Consequently, the corresponding couplings enter $\mathcal{H}_0$ without an additional factor of $1/2$.

\subsubsection{Finite second-neighbor exchange $J_2$ with $J_{3b}=0$}
\label{sec:j2_hourglass}

\begin{figure}[tb]
	\centering
	\includegraphics[width=0.9\linewidth]{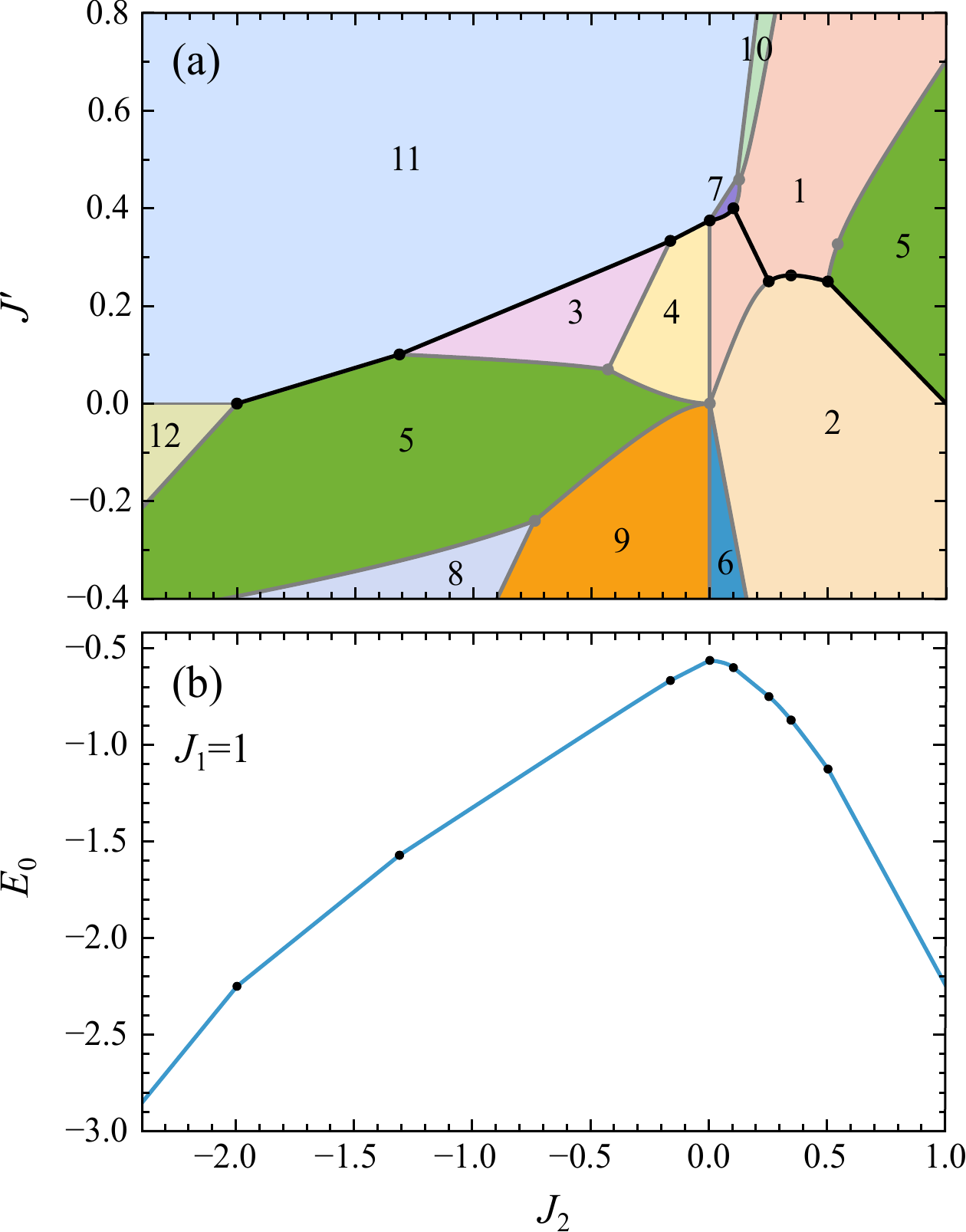}
	\caption{(a) Ground-state symmetry sectors of the hourglass Hamiltonian with second-neighbor coupling $J_2$, as a function of $J'$. 
	  Grey lines denote transitions between distinct symmetry sectors. The black curve marks the loci  of maximal ground-state energies across symmetry sectors, which define a lower bound on the true ground-state energy.
		(b) Lower bound $E_0$ for different values of $J_2$. The lower bound closely follows the boundaries between regions where the ground state becomes degenerate.}
	\label{fig:2_nn_states}
\end{figure}

We first consider the case $J_{3b}=0$, where the hourglass Hamiltonian includes nearest- and second-neighbor couplings only.
To determine the ground-state energy $E_0$ as a function of the parameters $J_2$ and $J'$, we apply the same symmetry-based direct diagonalization procedure introduced in Sec.~\ref{sec:direct_diag_hourglass} for the nearest-neighbor model: we block-diagonalize the hourglass Hamiltonian $\mathcal{H}_0$ according to the irreducible representations of the $D_{3d}$ point group (see Table~\ref{tab:D3d_char}).

The resulting energy spectrum as a function of $J_2$ and $J'$ is summarized in Table~\ref{tab:j3_0_spectrum}.
From this spectrum, we identify the ground state as the lowest-energy state within each symmetry sector.
Figure~\ref{fig:2_nn_states}(a) shows how the symmetry of the ground state evolves as a function of $J'$ for representative values of $J_2$, with grey lines marking transitions between distinct symmetry sectors.
To obtain a rigorous lower bound, we maximize the ground-state energy over $J'$ at fixed $J_2$.
The resulting optimal energies are shown in Fig.~\ref{fig:2_nn_states}(b).

In most parameter regions, the maximal energy occurs at points where two symmetry-distinct energy levels become degenerate, leading to piecewise analytic expressions for the bound.
An exception arises in the ferromagnetic regime, where the ground-state energy is independent of $J'$, and the optimization becomes trivial.

Overall, the inclusion of a finite second-neighbor coupling $J_2$ enriches the structure of the hourglass spectrum but does not alter the basic mechanism by which the optimal lower bound is determined:
Level crossings between symmetry-resolved energy branches of the hourglass Hamiltonian set the bound.

\subsubsection{A special point: $J_2=J_{3b}$}

We focus on the symmetric case
\begin{equation}
	J_2 = J_{3b} \equiv J_{23},
\end{equation}
which restores the full $(\mathfrak{S}_3 \times \mathfrak{S}_3)\rtimes \mathbb{Z}_2$ symmetry defined in Eq.~(\ref{eq:S3S3Z2}), a prerequisite for the analytic construction of Sec.~\ref{sec:analytic_hourglass}.
In this regime, we rewrite the hourglass Hamiltonian~(\ref{eq:motifham7_J1J2J3}) entirely in terms of total-spin operators as
\begin{multline}
	\mathcal{H}_0(J_1,J_{23},J^\prime)
	=
	\mathcal{H}_0(J_1,J^\prime)
	+
	J_{23}\,
	\mathbf{S}_{234}\cdot\mathbf{S}_{567},
	\label{eq:motifham7_J23}
\end{multline}
where $\mathbf{S}_{234}=\mathbf{S}_2+\mathbf{S}_3+\mathbf{S}_4$ and
$\mathbf{S}_{567}=\mathbf{S}_5+\mathbf{S}_6+\mathbf{S}_7$, as defined in Eqs.~(\ref{eq:s234etc}).
The Hamiltonian depends linearly on the coupling constants and can be expressed entirely in terms of total-spin operators as
\begin{multline}
	\mathcal{H}_0 =
	\frac{J^\prime}{2} \mathbf{S}_\text{tot}^2	+ 
	\left(\frac{J_{23}}{2} - \frac{J^\prime}{2}\right) \mathbf{S}_{234567}^2
	\\
	+\left(\frac{J_1}{4}-\frac{J^\prime}{2} -\frac{J_{23}}{2}\right)
	\left[\mathbf{S}_{234}^2+\mathbf{S}_{567}^2\right]
	\\
	-\left(\frac{3 J_1}{2}-\frac{5 J^\prime}{2} \right)S(S+1).
	\label{eq:hourglass_J23_totalspin}
\end{multline}
which provides analytic expressions for the energy eigenvalues and avoids explicit diagonalization of large Hamiltonian matrices.

In contrast to the nearest-neighbor case, the two basal triangles $(234)$ and $(567)$ are now directly coupled by $J_{23}$. As a result, the Hamiltonian is no longer bipartite in the sense required by the Lieb--Mattis theorem~\cite{Lieb}, and the ordering of total-spin multiplets cannot be inferred \emph{a priori}.
Nevertheless, for fixed quantum numbers
$(S_{234},S_{567},S_{234567},S_{\mathrm{tot}})$
the energy is an explicit linear function of the couplings $J_1$, $J_{23}$, and $J^\prime$.
We generate all allowed sets of spin quantum numbers
$(S_{234}, S_{567}, S_{234567}, S_{\mathrm{tot}})$
by systematically applying SU(2) spin-addition rules to the constituent spins of the hourglass motif.
For any fixed $J_1$ and $J_{23}$, we get the optimal lower bound by maximizing the minimum of this finite set of linear functions with respect to $J^\prime$. 
This constitutes a simple linear-optimization problem that we repeat as we scan the $(J_1, J_{23})$ parameter space.

The resulting lower bounds form piecewise linear functions of $J_1$ and $J_{23}$.
Changes in slope correspond to level crossings between distinct spin sectors of the hourglass Hamiltonian.
For $S=\tfrac12$, we summarize the complete set of bounds and the corresponding optimal quantum numbers in Table~\ref{tab:lower_Bound_J23}.
The same construction applies for $S\geq1$, although the number of relevant spin sectors increases rapidly.

While the analytic structure of the bound permits an explicit identification of the spin sectors that determine the optimum in each parameter region, the resulting pattern is generally nontrivial.
In particular, the dominant sector can change repeatedly as a function of $J_{23}$, reflecting the proliferation of level crossings inherent to the problem.
Nevertheless, for sufficiently small positive or negative values of $J_{23}$, the lowest-energy states coincide with those that optimize the $J_{23}=0$ problem and obey Lieb--Mattis ordering.

More specifically, for ferromagnetic $J_{23}<0$, the optimal solution arises from the level crossing between the
$(S_{234},S_{567},S_{234567},S_\mathrm{tot})=(S,S,2S,S)$
and
$(S+1,S+1,2S+2,S+2)$
spin sectors, leading to the lower bound
\begin{equation}
e_\mathrm{LB}(J_1,J_{23}) =
- S(S+1)
\left(
\frac{4S+1}{4S+2} J_1
-
\frac{2S+1}{2S+2} J_{23}
\right),
\end{equation}
valid in the interval
\begin{equation}
-\frac{|S-1|+1}{2S+1} J_1
\;\leq\;
J_{23}
\;\leq\;
0.
\end{equation}

For antiferromagnetic $J_{23}>0$, the lowest-energy states originate from the
$(S,S,2S,S)$,
$(S+1,S,2S+1,S+1)$, and
$(S,S+1,2S+1,S+1)$
sectors, leading to
\begin{equation}
e_\mathrm{LB}(J_1,J_{23})  =
- S(S+1)
\left(
\frac{4S+1}{4S+2} J_1
-
\frac{2S}{2S+1} J_{23}
\right),
\end{equation}
which holds for
\begin{equation}
0
\;\leq\;
J_{23}
\;\leq\;
\frac{S}{6S+2} J_1 .
\end{equation}
Both intervals remain finite in the classical limit $S\to\infty$.

\subsection{Bilinear and quadrilinear (ring exchange) spin-1/2 Hamiltonian}

We now extend the nearest-neighbor Heisenberg model on the pyrochlore lattice by including quadrilinear exchange terms of the form
$(\mathbf{S}_i\cdot\mathbf{S}_j)(\mathbf{S}_k\cdot\mathbf{S}_l)$.
Such interactions naturally arise as ring-exchange processes and were previously discussed in the context of the Klein point~\cite{nussinov2007}.
The corresponding lattice Hamiltonian reads
\begin{equation}
	\mathcal{H}
	=
	J \sum_{\langle ij \rangle\in\plaquette} \mathbf{S}_i\cdot \mathbf{S}_j
	+
	K \sum_{\langle ijkl \rangle\in\plaquette}
	(\mathbf{S}_i \cdot \mathbf{S}_j)(\mathbf{S}_k \cdot \mathbf{S}_l),
	\label{eq:pyroham_4site}
\end{equation}
where both sums run over individual tetrahedra.
For a tetrahedron with sites $(1,2,3,4)$, the quadrilinear term includes the three pairings
$(1,2)(3,4)$, $(1,3)(2,4)$, and $(1,4)(2,3)$.

At the special coupling ratio
\begin{equation}
	K=\frac{4}{5}J,
\end{equation}
the Hamiltonian becomes a sum of local projectors,
\begin{equation}
	\mathcal{H}_K
	=
	\frac{12}{5}J
	\sum_{\plaquette}\mathcal{P}_{\plaquette},
\end{equation}
where $\mathcal{P}_{\plaquette}$ projects onto the total-spin $S_{\plaquette}=2$ subspace of a tetrahedron.
Explicitly,
\begin{equation}
	\mathcal{P}_{\plaquette}
	=
	-\frac{1}{12}\mathbf{S}_{\plaquette}^2
	+
	\frac{1}{24}\mathbf{S}_{\plaquette}^4,
\end{equation}
with $\mathbf{S}_{\plaquette}=\mathbf{S}_i+\mathbf{S}_j+\mathbf{S}_k+\mathbf{S}_l$.
At this Klein point, the extensively degenerate ground-state manifold consists of singlet coverings with one singlet per tetrahedron, and the corresponding ground-state energy per site is
\begin{equation}
	\frac{E_\mathrm{GS}}{N}=-\frac{3}{8}J.
\end{equation}

To assess this model within Anderson’s lower-bound construction, we extend the hourglass sub-Hamiltonian~(\ref{eq:motifham7_nn}) by including the corresponding quadrilinear interaction,
\begin{align}
	\mathcal{H}_0(J,K,J^\prime)
	=
	\mathcal{H}_0(J,J^\prime)
	+
	\frac{K}{4}
	\sum_{\langle ijkl \rangle}
	(\mathbf{S}_i \cdot \mathbf{S}_j)
	(\mathbf{S}_k \cdot \mathbf{S}_l),
	\label{eq:motifham7_4exchange}
\end{align}
where the sum runs over the quadrilinear terms contained within the hourglass motif.
We determine the lower bound on the ground-state energy as a function of $K$ by applying the same symmetry-based diagonalization procedure used throughout this work (see Sec.~\ref{sec:direct_diag_hourglass}). 

\begin{figure}[tb]
	\centering
	\includegraphics[width=0.9\linewidth]{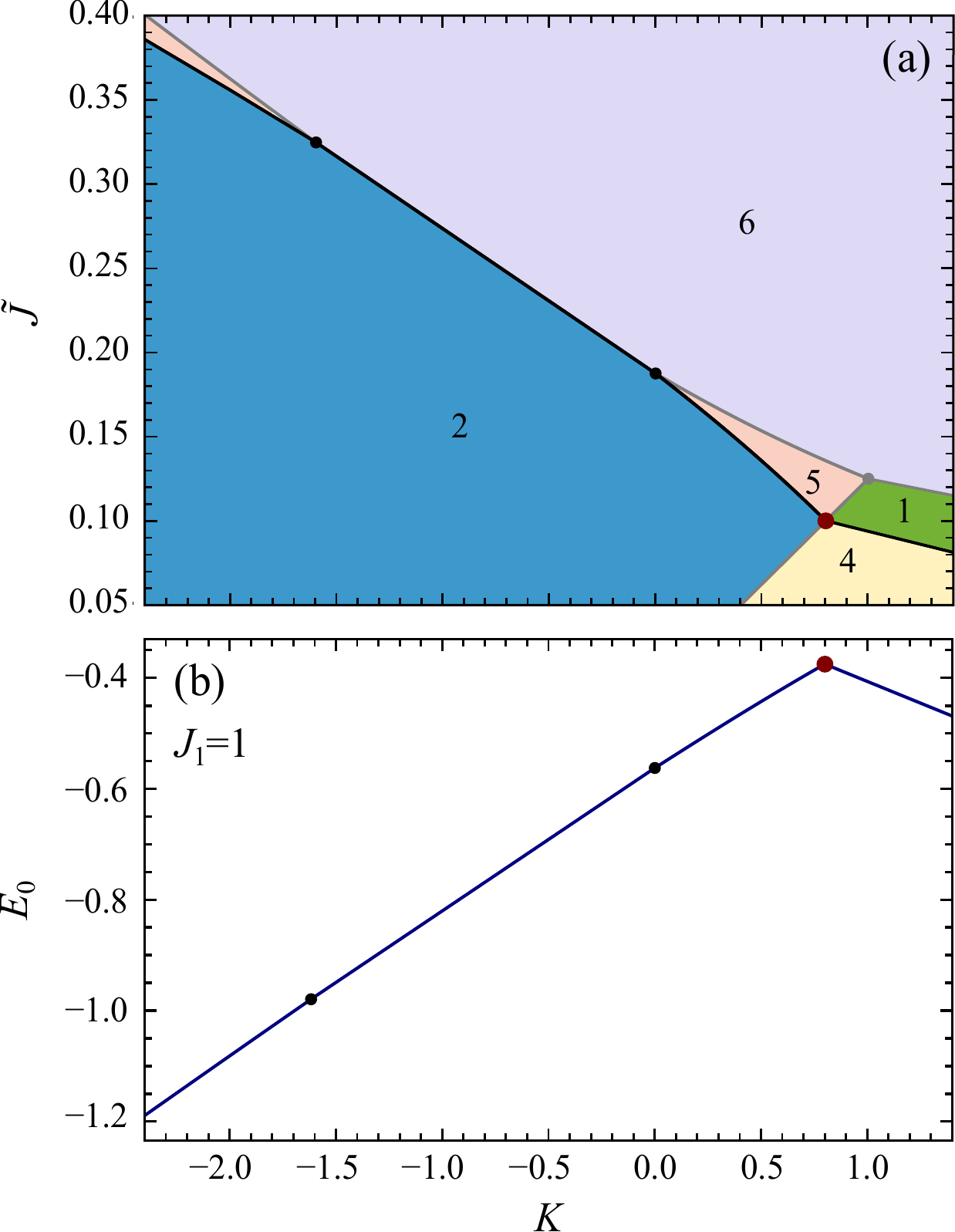}
	\caption{(a) Ground-state symmetry sectors of the hourglass Hamiltonian with bilinear and quadrilinear couplings, as a function of $\tilde{J}$, which is defined in  Sec.~\ref{sec:foursite_appendix}.  Grey lines denote transitions between distinct symmetry sectors. The black curve marks the loci of maximal ground-state energies across symmetry sectors, which define a lower bound on the true ground-state energy. The burgundy dot marks the Klein point for $K/J=4/5$. 
	(b) Lower bound $E_0$ for different values of $K$.
 The lower bound closely follows the boundaries between regions where the ground state becomes degenerate.  The burgundy dot marks the Klein point.}
	\label{fig:4ex_energies_states}
\end{figure}

As in the bilinear cases discussed above, we first analyze the symmetry-resolved energy landscape of the sub-Hamiltonian as a function of the free parameter. 
We derive closed-form expressions for all symmetry-resolved energy levels as functions of the free parameter and the ring-exchange coupling.
For clarity, we omit these expressions from the main text and present the explicit form of the hourglass Hamiltonian in Eq.~(\ref{eq:ringexchangewithP}), together with the complete set of symmetry-resolved energy levels in Table~\ref{tab:4site_exchange_energies_explicit}, in Appendix~\ref{sec:foursite_appendix}.
We then use these analytic energies to construct the energy landscape as a function of \(\tilde{J}\), defined in Eq.~(\ref{eq:Jtildedef}) and shown in Fig.~\ref{fig:4ex_energies_states}(a).

For fixed $K$, the ground state belongs to one of these symmetry sectors, and grey lines indicate transitions between distinct sectors as $\tilde{J}$ is varied.
The lower bound follows from the envelope of these lowest energies.
Figure~\ref{fig:4ex_energies_states}(b) shows the resulting bound $E_0(K)$, obtained by maximizing, with respect to the free parameter $\tilde J$, the minimum energy among all symmetry sectors.
As in the $J_2$ and $J_{23}$ cases, the optimal bound typically occurs at points where two symmetry-resolved energy levels become degenerate.

At the Klein point (marked by the burgundy dot in Fig.~\ref{fig:4ex_energies_states}(b)), the lower bound coincides with the variational ground-state energy obtained from the exact singlet-covering construction, indicating that Anderson’s bound saturates in this case. This agreement provides a consistency check of the lower-bound approach in the presence of quadrilinear interactions.

\subsection{Chiral interaction}
\label{sec:chiral_interaction}

We now consider the scalar spin-chirality interaction on the pyrochlore lattice.
Motivated by the classical analysis of Ref.~\cite{Lozano2024}, we study the quantum Hamiltonian built from the scalar chirality on an oriented triangle $(i,j,k)$,
\begin{equation}
	\hat \chi_{ijk}
	=
	\mathbf{S}_i \cdot \left( \mathbf{S}_j \times \mathbf{S}_k \right).
\end{equation}
On a tetrahedron with sites $(1,2,3,4)$, we define the total chirality as
\begin{equation}
	\hat \chi_{\plaquette}
	=
	\hat \chi_{123}
	+
	\hat \chi_{142}
	+
	\hat \chi_{134}
	+
	\hat \chi_{243},
	\label{eq:tetrahedronchirality}
\end{equation}
which fixes a consistent orientation of the four triangular faces of a single tetrahedron.

Using the tetrahedral chirality $\hat \chi_{\plaquette}$, we consider two distinct lattice patterns.

\medskip
\noindent
\emph{(i) Uniform chirality.}
\begin{equation}
	\mathcal{H}_{\chi}^{\mathrm{uni}}
	=
	J_\chi \sum_{\plaquette} \hat \chi_{\plaquette},
	\label{eq:pyroham_chiral_uniform}
\end{equation}
where all tetrahedra enter with the same sign, resulting in an identical orientation of their triangular loops when viewed from the interior.
Because the scalar chirality
$\hat \chi_{ijk}=\mathbf{S}_i\cdot(\mathbf{S}_j\times\mathbf{S}_k)$
changes sign under any orientation-reversing spatial operation, the uniform chiral interaction is invariant only under proper, orientation-preserving lattice symmetries.
In particular, inversion and mirror symmetries are explicitly broken, while $C_2$ twofold rotations are preserved.
Time-reversal symmetry is also broken, since $\mathcal{T}:\hat \chi_{ijk}\rightarrow -\hat \chi_{ijk}$.
As a consequence, the uniform chiral Hamiltonian preserves neither inversion, mirror, nor time reversal individually, but only their combination.

Taking these into account, the space group symmetry of the Hamiltonian is reduced from the full octahedral group $O_h$ to the chiral cubic group $O$, corresponding to the space-group subgroup ($S_\mathrm{uni}$) $F4_132$ (No.~210)~\cite{ITC_A}.

\medskip
\noindent
\emph{(ii) Alternating chirality.}
\begin{equation}
	\mathcal{H}_{\chi}^{\mathrm{alt}}
	=
	J_\chi \sum_{\plaquette} \eta_{\plaquette}\,\hat \chi_{\plaquette},
	\qquad
	\eta_{\plaquette}=\pm 1,
	\label{eq:pyroham_chiral_alternating}
\end{equation}
where the $\eta_{\plaquette}$ encodes a staggered sign pattern and alternates between tetrahedra related by inversion (\emph{i.e.}, between the two tetrahedral sublattices).
Under inversion, both the orientation of the triangular loops and the sign factor $\eta_{\plaquette}$ change, leaving $\mathcal{H}_{\chi}^{\mathrm{alt}}$ invariant.
Thus, the alternating chiral pattern restores inversion symmetry at the lattice level, while breaking time-reversal, $C_2$ rotation, and mirror symmetry.

Taking these into account, the space group symmetry of the Hamiltonian is reduced from the full octahedral group $O_h$ to the chiral cubic group $T_h$, corresponding to the space-group subgroup ($S_\mathrm{alt}$) $Fd\bar{3}$ (No. 203)~\cite{ITC_A}.

Scalar chiral interactions have been studied extensively on the kagome lattice, both in classical and quantum settings \cite{Pitts2022, Fabrizio2022}.
By contrast, the quantum pyrochlore case remains largely unexplored.
Here, we do not attempt a systematic study; instead, we use Anderson’s construction to obtain a lower bound on the ground-state energy of the spin-$\tfrac12$ model.

\begin{figure}[tb]
	\centering
	\includegraphics[width=0.9\linewidth]{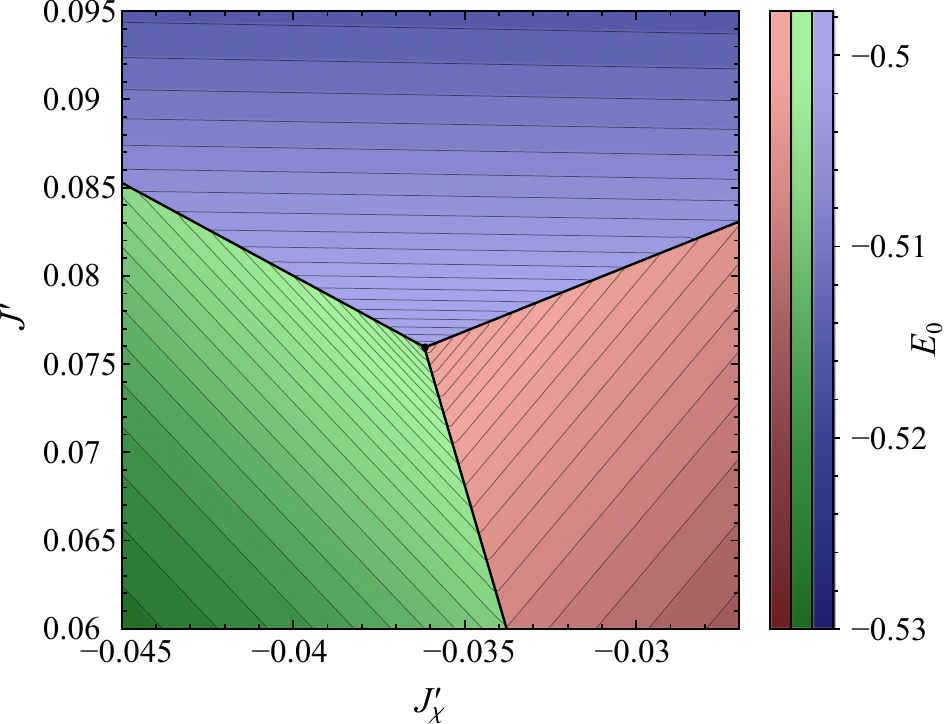}
	\caption{Ground-state symmetry sectors in the vicinity of the maximal ground-state energy, indicated by the black dot. The three relevant energy levels, shown in red, green, and blue, are listed in Eq.~(\ref{eq:chiral_ene_red},\ref{eq:chiral_ene_green},\ref{eq:chiral_ene_blue}) respectively. The shading illustrates how the ground-state energy varies as the parameters ($ J^\prime$, $ J_\chi^\prime$) are tuned away from their optimal values.}
	\label{fig:contour_chiral}
\end{figure}

 \subsubsection{Alternating chirality: hourglass lower bound}

In the alternating pattern, tetrahedra related by inversion carry opposite signs of the tetrahedral chirality.
On the hourglass motif, this alternating pattern assigns opposite chiral couplings to the two corner-sharing tetrahedra.
Consequently, only those orientation-reversing spatial symmetries that also exchange the two tetrahedra can remain symmetries of the Hamiltonian.
This requirement reduces the geometric point group $D_{3d}$ to its subgroup $C_{3i}$ generated by the threefold rotation $C_3$ about the hourglass axis and inversion $i$, which interchanges the two tetrahedra,
\begin{equation}
	C_{3i} \cong	C_3 \times C_i .
\end{equation}
The order of $C_{3i}$ is six,  and it is also denoted by $S_6$.

Within these symmetry constraints, the most general sub-Hamiltonian consistent with SU(2) spin symmetry, alternating chirality, and the hourglass geometry reads
\begin{align}
		\mathcal{H}_0(J_\chi,J^\prime,J^\prime_\chi)
		=&\,
		\mathcal{H}_0(0,J^\prime)
		-4J^\prime_\chi \left( \hat \chi_{243}+ \eta \hat \chi_{576} \right)
		\nonumber\\
		&+\left(\frac{4}{3}J^\prime_\chi-\frac{J_\chi}{3}\right)
		\left(
		\hat \chi_{123}+\hat \chi_{142}+\hat \chi_{134}
		\right)
		\nonumber\\
		&+\eta\left(\frac{4}{3}J^\prime_\chi-\frac{J_\chi}{3}\right)
		\left(
		\hat \chi_{156}+\hat \chi_{167}+\hat \chi_{175}
		\right),
		\label{eq:motifham7_chiral_eta}
\end{align}
where $\eta=-1$. Here $\mathcal{H}_0(0,J^\prime)$ denotes the nearest-neighbor hourglass Hamiltonian defined in Eq.~(\ref{eq:motifham7_nn}).
The lattice sum eliminates the free parameters $J^\prime$ and $J^\prime_\chi$.

To represent the Hamiltonian $\mathcal{H}_0$ in the computational basis, we exploit the fact that for $S=1/2$ the scalar chirality operator admits a representation in terms of three-site permutation operators,
\begin{equation}
	\hat \chi_{ijk}
	=
	\frac{1}{4 i}
	\left(
	\mathcal{P}_{ijk}
	-
	\mathcal{P}_{ikj}
	\right).
\end{equation}
We determine the lower bound by diagonalizing $\mathcal{H}_0$ in the symmetry-resolved Hilbert space of the hourglass motif and maximizing the resulting ground-state energy with respect to the free parameters $J^\prime$ and $J^\prime_\chi$, as shown in Fig.~\ref{fig:contour_chiral}.

For spin-$\tfrac12$, only three levels compete for the lowest energy near the optimum: there are two $S_\mathrm{tot}=\tfrac12$ levels, one (red in Fig.~\ref{fig:contour_chiral}) transforms according to the two-dimensional irrep $E_g$ and has energy
\begin{equation}
E_{1/2}^{(E_g)}
=
\frac{J^\prime}{2}
-
\frac{1}{\sqrt{3}}\left(1+2J^\prime_\chi\right),
\label{eq:chiral_ene_red}
\end{equation}
and the other (green in Fig.~\ref{fig:contour_chiral}) belongs to the one-dimensional $A_g$ and $A_u$ irreducible representations and has energy
\begin{equation}
E_{1/2}^{(A)} =
J^\prime - \frac{1}{2} \sqrt{(J^\prime)^2 + (1-4J^\prime_\chi)^2}.
\label{eq:chiral_ene_green}
\end{equation}
The lowest energy state in the $S_\mathrm{tot}=\tfrac32$ sector transforms according to the $E_g$ and $E_u$ irreps and has energy
\begin{widetext}
\begin{equation}
E_{3/2} =
-\frac{J^\prime}{2}
- \frac{1}{4\sqrt{3}} 
 \left[
    1 + 2\sqrt{ 12(J^\prime)^2 + \sqrt{3}J^\prime(1-4J^\prime_\chi) + (1-4J^\prime_\chi)^2} + 8J^\prime_\chi
\right],
\label{eq:chiral_ene_blue}
\end{equation}
\end{widetext}
shown in blue.

The maximal lower bound arises at the unique point in parameter space where these three levels become degenerate. This triple crossing fixes the optimal values of the free parameters 
\begin{subequations}
\begin{align}
	J^\prime & = \frac{\sqrt{123}-5\sqrt{3}}{32} \approx 0.075946,
\\
	J^\prime_\chi &= \frac{53-9\sqrt{41}}{128} \approx -0.036157,
\end{align}
\end{subequations}
resulting in the lower bound
\begin{equation}
	e_\mathrm{LB}
	= \frac{\sqrt{3}(\sqrt{41}-11)}{16}
	\approx -0.497626(3).
	\label{eq:chiral_LB_hourglass}
\end{equation}

For comparison, setting $J^\prime=0$ leads to a substantially weaker bound.
In that case, the optimal solution occurs at
$J^\prime_\chi=\tfrac{1}{8}(5-3\sqrt{3})$,
with
\begin{equation}
	e_\mathrm{LB}
	=
	-\frac{3}{4} \left(\sqrt{3}-1\right)
	=
	-0.549038(1).
\end{equation}
This comparison demonstrates that allowing both free parameters is essential for capturing the energetics of the alternating chiral interaction.
By contrast, applying the same procedure to the purely Heisenberg hourglass Hamiltonian~(\ref{eq:motifham7_nn}) leaves the optimal solution at $J^\prime_\chi=0$, restoring the higher $D_{3d}$ symmetry.

Finally, we note that the effective symmetry of the hourglass Hamiltonian is larger than the geometric point group $C_{3i}$.
The upper and lower basal triangles can be rotated independently, generating a $C_3 \times C_3$ symmetry, while inversion exchanges the two triangles and provides an additional $\mathbb{Z}_2$ operation.
Taken together, these transformations generate a group isomorphic to
\begin{equation}
	(C_3 \times C_3)\rtimes \mathbb{Z}_2, 
	\label{eq:groupC3C3Z2}
\end{equation}
a group of order $18$.

 \subsubsection{Uniform chirality: hourglass lower bound}
We now repeat the lower-bound construction for the Hamiltonian with uniform scalar chirality, Eq.~(\ref{eq:pyroham_chiral_uniform}), for which all tetrahedra carry the same orientation of triangular loops. 
%
 Within the hourglass motif, the two corner-sharing tetrahedra are related by a proper $C_2'$ rotation, which --using the site labeling of Fig.~\ref{fig:hourglass_star_tatra}(c)-- is implemented by the  permutation
$(3\ 6)(2\ 5)(4\ 7)$. 
We exclude all improper elements of the geometric point group $D_{3d}$, including inversion and mirror reflections.
The remaining symmetry operations form  
\begin{equation}
	D_3 \cong C_3 \times C_2,
\end{equation}
which has 6 elements.

The Hamiltonian becomes equivalent to
Eq.~(\ref{eq:motifham7_chiral_eta}) with $\eta=-1$ by relabeling the sites within a tetrahedron (e.g. $5 \leftrightarrow 6$), which is allowed
in the absence of longer-range exchange interactions. Thus, the lower bound is the same as for the alternating case, Eq.~(\ref{eq:chiral_LB_hourglass}).

\subsubsection{Comparison with variational upper bounds}

We now benchmark the rigorous hourglass lower bound against simple variational upper bounds to bracket the true ground-state energy.

We first consider a classical variational ansatz constructed from the chiral tetrahedral states of Ref.~\onlinecite{Lozano2024}, implemented as a product of spin–coherent states.
This approximation treats the spins as classical vectors and neglects entanglement.
It yields the variational energy per site
\begin{equation}
	\frac{E_0}{N}
	=
	-\frac{2\sqrt{3}}{9}
	\approx
	-0.384900(1).
\end{equation}

A lower variational energy is achieved when we allow entanglement within individual tetrahedra.
In the spin-singlet sector, the scalar chiral interaction on a single tetrahedron admits two chiral eigenstates,
\begin{multline}
  \ket{\chi^\pm}=
   \ket{\uparrow\uparrow\downarrow\downarrow}
   +\ket{\downarrow\downarrow\uparrow\uparrow}
   \\ 
+e^{\pm i2\pi/3}
\left(\ket{\downarrow\uparrow\uparrow\downarrow}
+\ket{\uparrow\downarrow\downarrow\uparrow}\right)
\\
+e^{\mp i2\pi/3}
\left(\ket{\uparrow\downarrow\uparrow\downarrow}
+\ket{\downarrow\uparrow\downarrow\uparrow}\right),
\end{multline}
which satisfy
\begin{equation}
	\hat \chi_{\plaquette}\ket{\chi^\pm}
	=
	\pm\sqrt{3}\ket{\chi^\pm}.
\end{equation}
Depending on the sign of the chiral coupling, one of these states minimizes the energy of an isolated tetrahedron.

We make use of the bipartite tiling of the pyrochlore lattice by tetrahedra and construct a variational wave function as a direct product of such four-spin chiral ground states on one tetrahedral sublattice.
Evaluating either the uniform chiral Hamiltonian Eq.~(\ref{eq:pyroham_chiral_uniform})
or the alternating chiral Hamiltonian  Eq.~(\ref{eq:pyroham_chiral_alternating})  this variational state gives the same upper bound,
\begin{equation}
  \frac{E_{\mathrm{GS}}}{N}
  \leq 
  -\frac{\sqrt{3}}{4}
  \approx
  -0.433012(7) .
\end{equation}
Although the uniform and alternating chiral Hamiltonians differ in their assignment of chirality signs to tetrahedra, this distinction is not resolved by the present variational state. 
Since the wave function is a direct product of four-spin spin-singlet chiral states on a single tetrahedral sublattice, scalar chirality operators acting on tetrahedra of the complementary sublattice have vanishing expectation values. As a result, only tetrahedra hosting the local four-spin chiral states contribute to the energy, rendering the variational estimate insensitive to whether the chirality pattern is uniform or alternating.

 Following Sec.~\ref{sec:tetra}, we can calculate the lower bound based on a single tetrahedron:
\begin{equation}
	e_\mathrm{LB}=-\frac{\sqrt{3}}{2} \approx -0.866025.
\end{equation}
This bound is substantially weaker than the one derived from the two-tetrahedra hourglass motif.

Combining the strongest variational upper bound with the rigorous hourglass lower bound, we arrive at  the inequality
\begin{equation}
 - \frac{\sqrt{3}(11-\sqrt{41})}{16} 
  \le
 \frac{E_{\mathrm{GS}}}{N}
 \;\le\;
  -\frac{\sqrt{3}}{4}
\end{equation}
or, numerically, 
\begin{equation}
 -0.497626
 \;\lessapprox \;
 \frac{E_{\mathrm{GS}}}{N}
 \;\lessapprox \;
 -0.433012 
\end{equation}
which applies to both the uniform and alternating chiral interactions.

\section{Beyond 7-sites}
\label{sec:crown}

Recent numerical studies indicate that resonating singlet states on hexagonal loops play an essential role in the low-energy physics of the pyrochlore antiferromagnet \cite{schafer}.
 Motivated by these findings, we extend our studies to larger clusters that explicitly contain such hexagonal motifs, notably to the 18-site cluster shown in Fig.~\ref{fig:crown}.  
 
\subsection{Hexagons and the 18-site crown cluster}

As a first step, it is instructive to consider six-site hexagonal rings.
We build the pyrochlore lattice Heisenberg Hamiltonian as a sum of hexagons in Eq.~(\ref{eq:lowb}).
The number of hexagons equals the number of lattice sites, and two hexagons share a nearest-neighbor bond. Consequently, the effective exchange within each hexagon is $J/2$.
The exact ground-state energy of a Heisenberg ring of six $S=\tfrac12$ spins with unit exchange coupling is
\begin{equation}
  E_{\mathrm{hex}} = -1-\sqrt{13}/2 \approx -2.803.
\end{equation}
leading to  the lower bound
\begin{equation}
 e_\mathrm{LB} = \frac{E_{\mathrm{hex}}}{2}  \approx -1.401
\end{equation}
This bound is rather weak, even compared to the simple Anderson star construction (Table~\ref{tab:bounds_comparison_transposed}).

Interestingly, when hexagons are instead used as variational building blocks—by covering the lattice with $N/6$ non-overlapping hexagons—the resulting variational energy
\begin{equation}
  \frac{E_{\mathrm{var}}}{N} = \frac{E_{\mathrm{hex}}}{6} \approx -0.467
\end{equation}
is surprisingly competitive.
This substantial mismatch highlights the limitations of small clusters in establishing rigorous lower bounds and indicates that larger motifs capturing both hexagonal loops and tetrahedral connectivity are required to obtain tighter bounds.

\begin{figure}[tb]
	\centering
	\includegraphics[width=0.6\linewidth]{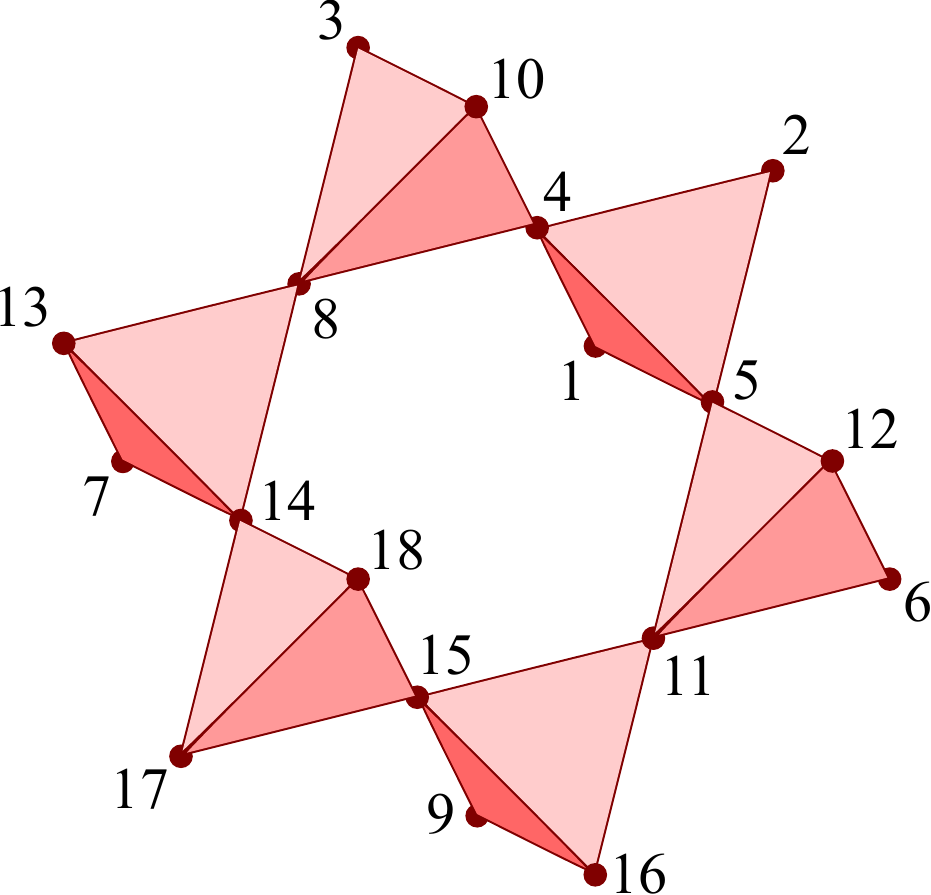}
	\caption{The 18-site ``crown'' motif of the pyrochlore lattice, consisting of six corner-sharing tetrahedra arranged in a ring.}
	\label{fig:crown}
\end{figure}

Guided by this insight, we turn to the 18-site “crown” cluster shown in Fig.~\ref{fig:crown}.
This motif represents the minimal pyrochlore subgraph that simultaneously (i) contains a closed hexagonal loop, (ii) preserves full tetrahedral connectivity around each bond of the hexagon, and (iii) embeds the seven-site hourglass motif as a substructure.
As such, it provides a natural extension of previously studied clusters, incorporating hexagonal loop correlations and tetrahedral frustration simultaneously.
The most general bilinear spin Hamiltonian compatible with the symmetry of the crown cluster contains 19 symmetry-inequivalent bond orbits. Imposing the constraints required by embedding the motif into the pyrochlore lattice reduces this number to 9 independent parameters.
The explicit Hamiltonian and the corresponding embedding constraints are given in Appendix~\ref{sec:appendix_crown}, Eq.~(\ref{eq:crown_hamiltonian}).

\subsection{Spin-$\tfrac12$ crown: brute-force optimization}
\label{sec:crown_1o2}

For $S=\tfrac12$, the Hilbert space of the 18-site cluster in the $S^z_{\mathrm{tot}}=0$ sector has dimension $48\,620$, which allows for direct numerical treatment.
We construct the Hamiltonian matrix in this sector and determine its ground-state energy using Lánczos diagonalization for fixed values of the exchange parameters.
The resulting ground-state energy is then maximized over the parameter space using the Nelder–Mead simplex algorithm.

This procedure yields the lower bound
\begin{equation}
e_\mathrm{LB} = -0.549832(8).
\end{equation}
for the free parameters given in Eqs.~(\ref{eq:crown_cs}). It already represents a significant improvement over the seven-site hourglass bound.

\begin{figure}[tb]
	\centering
	\includegraphics[width=0.5\linewidth]{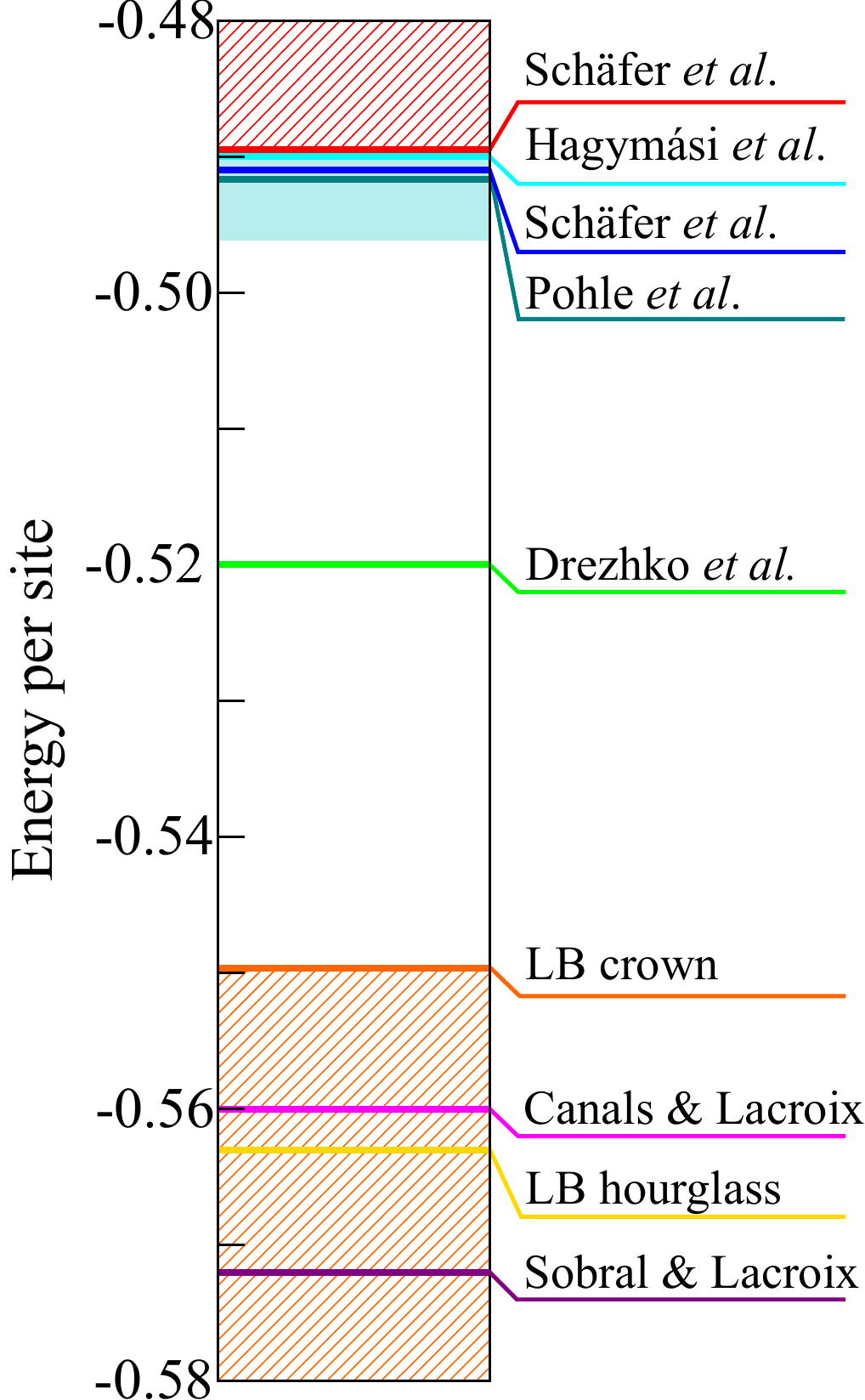}
	\caption{Various estimates for the $E_\mathrm{GS}/N$ ground-state energy per site (indicated by horizontal lines)  of the spin-1/2 antiferromagnetic Heisenberg model on the pyrochlore lattice in the thermodynamic limit with $J=1$: Sobral and Lacroix -0.572 \cite{sobral}, Canals and Lacroix -0.56 \cite{canals}, Drezhko \textit{et al.} -0.52 \cite{drezhko},  Schäfer \textit{et al.} -0.4917 \cite{schafer}, Hagymási \textit{et al.} -0.490$\pm$0.006 \cite{hagymasi}, Pohle \textit{et al.} -0.4921(4) \cite{pohle2023}. An upper bound obtained by Schäfer \textit{et al.} -0.489472 \cite{schafer}  and a lower bound obtained in this work -0.549832(8). }
	\label{fig:energies_s_1_2}
\end{figure}

Figure~\ref{fig:energies_s_1_2} compares this lower bound with available variational and numerical estimates for the ground-state energy of the spin-$\tfrac12$ pyrochlore antiferromagnet.
While a substantial gap remains, the crown construction provides one of the tightest rigorous lower bounds currently available.

\subsection{Reducing the Hilbert space: strategy for higher spins}

For $S=1$, a brute-force treatment becomes impractical: the dimension of the $S^z_{\mathrm{tot}}=0$ sector grows to $44\,152\,809$, beyond the capabilities of standard computational resources, since we need not only to diagonalize the Hamiltonian but also to optimize the free parameters. 
To make progress, we make use of additional conserved quantities suggested by the structure of the crown cluster.

For uniform nearest-neighbor exchange couplings within each tetrahedron, the Hamiltonian of the crown cluster conserves the total spin of the two sites that are not shared with neighboring tetrahedra. In this case, the squared total-spin operator of each outer pair,
\(
\tilde{\mathbf S}_{ij}^2 = (\mathbf S_i + \mathbf S_j)^2,
\)
commutes with the cluster Hamiltonian, following ideas put forward in Refs.~ \cite{Tarrach90,Xian_PRB.52.12485_couple_chains,Honecker_Mila_Troyer_2000,Lamas_Matera_PRB_92_Dimerized}. The spin of the pairs, $\tilde S_{ij}$, ceases to be a good quantum number once inequivalent couplings connect the outer spins to the rest of the cluster, as such terms mix different $\tilde S_{ij}$ sectors.

Motivated by this observation, we impose the commutation relations
\begin{equation}
	[\mathcal H_0, \tilde{\mathbf S}_{ij}^2] = 0,
\end{equation}
for the six outer spin pairs
\(
(i,j) = (1,2),(6,12),(9,16),(17,18),(7,13),(3,10).
\)
Evaluating these commutators explicitly using the Hamiltonian (\ref{eq:crown_hamiltonian}) yields the following additional constraints among the exchange parameters:
\begin{subequations}
\begin{align}
	c_2 &= c_6 = c_{11} = -2c_5, \\
	c_4 &= c_7 = 0, \\
	c_{12} &= c_3 .
\end{align}
\end{subequations}
With these constraints, only three free parameters remain: $c_1$, $c_3$, and $c_5$. In the Hamiltonian (\ref{eq:crown_red}) we keep only  $c_1$ and $c_3$, which correspond to nearest-neighbor exchanges, and we englect the second-neighbor $c_5$. 
Within this reduced description, the problem maps onto an effective hexagon formed by the inner spins (sites $4,5,11,15,14,8$), coupled to composite outer spin pairs that remain in sectors of fixed total spin.

\subsection{Results for $S=\tfrac12$ and $S=1$}
\label{sec:crown_reduced_1o2_1}

For $S=\tfrac12$, the lower bound
\begin{equation}
  e_\mathrm{LB} = -0.550158(2)
\end{equation}
is obtained for $c_1=0.038141(4)$ and $c_3=0.081795(7)$ at a level crossing between two distinct configurations of the outer spin pairs:
(i) alternating singlet and triplet pairs as we go around the hexagon, $[0,1,0,1,0,1]$, and
(ii) two singlets with the remaining pairs forming triplets, $[1,1,0,1,1,0]$.
The close agreement with the unrestricted optimization, including the location of this level crossing, confirms the robustness of the reduction scheme.

Applying the same strategy to $S=1$, the system maps onto a spin-1 hexagon coupled to composite outer pairs with total spin $0$, $1$, or $2$.
The reduced Hilbert space decomposes into sectors labeled by the outer-pair spin configuration.
Among these, the maximal block dimension in the $S^z_{\mathrm{tot}}=0$ sector is $1\,116\,220$, corresponding to the case where all outer pairs carry total spin $2$, and representing a reduction by a factor of approximately $40$ compared to the whole space.
Searching across the different outer-pair spin configurations, the optimal lower bound is
\[
e_\mathrm{LB} = -1.632985(8),
\]
obtained at $c_1=0.053466(8)$ and $c_3=0.082726(1)$, with outer-spin configurations $[0,0,0,2,0,2]$ and $[0,2,2,0,2,2]$. This is remarkably close to the  variational upper bound  given in Ref.~\cite{schafer}, which rules out energies larger than $-1.490(1)$.

\subsection{Relation to other cluster constructions}

In addition to the seven-site hourglass and the 18-site crown clusters discussed in detail in this work, we have examined several alternative motifs with comparable or larger numbers of sites, including two distinct 16-site clusters.

The first one is the ``cactus'' cluster formed by attaching an additional tetrahedron to each vertex of a central tetrahedron. It gives
\begin{equation}
  e_{\mathrm{LB}} = -0.560048(7),
\end{equation}
which represents a modest numerical improvement over the hourglass result.

The second is the super-tetrahedron introduced in Ref.~\cite{pohle2023}, which contains four tetrahedra at the corners and hexagonal loops on its faces, with a very poor
\begin{equation}
  e_{\mathrm{LB}} = -0.681155(0).
\end{equation}
These results show that different choices of cluster motifs can lead to substantially different numerical lower bounds.
In particular, increasing the cluster size alone does not guarantee an improvement of the lower bound on the ground-state energy.

When a larger cluster contains a given motif as a subcluster with compatible embedding constraints, the corresponding Anderson-type lower bound is guaranteed to be at least as tight as the motif bound. %
This monotonicity follows directly from the arguments presented in the introduction, with the Hamiltonian of the larger cluster replacing $\mathcal{H}$ in Eq.~(\ref{eq:lowb}).

\subsection{Chiral interaction on the crown cluster}

We next study the uniform and alternating chiral interactions on the pyrochlore lattice using the 18-site crown cluster for \( S=\tfrac{1}{2} \), following Sec.~\ref{sec:chiral_interaction}. 
In the sub-Hamiltonian, we again keep bilinear terms and set \( J_1 = J_2 = J_3 = 0 \) in the Hamiltonian \( \mathcal{H}_0 \) of Eq.~(\ref{eq:crown_hamiltonian}). 
Chiral interactions are included only on the tetrahedra. 
The chiral sub-Hamiltonian \( \mathcal{H}_{\chi_0}^{\mathrm{(uni)}} \) considered here contains three independent parameters and is given in Eq.~(\ref{eq:crown_hamiltonian_chiral}) in Appendix~\ref{sec:appendix_crown}. 
The full sub-Hamiltonian is therefore \( \mathcal{H}_0 + \mathcal{H}_{\chi_0}^{\mathrm{(uni)}} \).

Using the same brute-force optimization procedure as in Sec.~\ref{sec:crown_1o2}, we obtain a lower bound on the ground state energy with  $J_\chi=1$:
\begin{equation}
	e_\mathrm{LB} = -0.481197(6),
\end{equation}
for the parameter values listed in Eq.~(\ref{eq:crown_cs_chiral}). In the case of the alternating Hamiltonian, when the sub-Hamiltonian is $\mathcal{H}_0+\mathcal{H}_{\chi_0}^{\mathrm{(alt)}}$, the lower bound for the parameters given by Eqs.~(\ref{eq:crown_cs_chiral_alt}) gives 
\begin{equation}
	e_\mathrm{LB} = -0.481441(7),
\end{equation}
slightly below the uniform case. 
In the end, with the crown cluster, we could narrow the energy gap on the ground state energy per site for the uniform Hamiltonian to
\begin{equation}
	-0.481197(6)
	\;\leq\;
	\frac{E_{\mathrm{GS}}}{N}
	\;\lessapprox \;
	-0.433012\; ,
\end{equation}
and for the alternating Hamiltonian
\begin{equation}
	-0.481441(7)
	\;\leq\;
	\frac{E_{\mathrm{GS}}}{N}
	\;\lessapprox \;
	-0.433012\; .
\end{equation}

\section{Conclusions}
\label{sec:conclusions}

In this work, we revisit Anderson’s method for deriving lower bounds on the ground-state energy and apply it to the Heisenberg model on the pyrochlore lattice. To this end, we systematically construct motifs of increasing size that are compatible with the space-group symmetries of the lattice. These motifs serve as a basis for defining trial Hamiltonians that, beyond the Heisenberg interactions, include additional terms absent from the original model. Although these additional terms vanish upon summation over the lattice, they can quantitatively improve the lower bounds.

Focusing first on the pyrochlore lattice with nearest-neighbor Heisenberg exchange, we begin with the simplest constructions: Anderson’s original star and the tetrahedron, the fundamental building block of the lattice. As expected, the star—optimized for bipartite systems—substantially underestimates the ground-state energy. The tetrahedron yields an appreciably improved bound, underscoring the importance of incorporating frustration at the level of the motif. This observation motivates the introduction of the hourglass motif, formed by two tetrahedra sharing a single site.

The seven-site hourglass constitutes the first nontrivial improvement, as it introduces a free parameter that controls the relative strengths of the two symmetry-inequivalent nearest-neighbor bonds within the motif. By optimizing this parameter, we further improve the lower bound for a single tetrahedron. Moreover, the simplicity and high symmetry of the trial Hamiltonian allow the lowest energy to be expressed in closed analytical form in terms of total-spin quantum numbers, also for spins larger than $S=1/2$.

Motivated by recent numerical evidence emphasizing the role of resonating hexagonal loops in the pyrochlore antiferromagnet \cite{schafer}, we further extend the construction to the 18-site crown cluster. This motif is the smallest subgraph that simultaneously contains a closed hexagonal loop, preserves full tetrahedral connectivity around each bond of the loop, and embeds the hourglass as a substructure. For $S=1/2$, a brute-force optimization over nine free parameters within this cluster yields the tightest rigorous lower bounds currently available. For $S=1$, we employ a reduction scheme based on conserved outer-pair spin quantum numbers to decrease the dimension of the Hilbert space by an order of magnitude. This reduction enables the computation of a lower bound that lies beyond the reach of the brute-force approach. A summary of our results for the nearest-neighbor Heisenberg model is presented in Table~\ref{tab:bounds_comparison_transposed}.

We further applied the Anderson-based construction to extensions of the nearest neighbor Heisenberg model, including second- and third-neighbor exchange interactions, four-spin ring exchange, and scalar spin-chirality terms. 
In this context, the hourglass motif serves as the smallest nontrivial cluster for constructing the lower bounds. 
In all cases, the resulting bounds are determined by level crossings between symmetry-distinct sectors, yielding piecewise-analytic expressions. 
At a particular parameter point—the Klein point of the bilinear–quadrilinear model—the Anderson bound is exactly saturated, providing a consistency check of the method.
For the chiral interaction, we additionally considered the 18-site crown cluster and, independently, constructed a variational upper bound, thereby bracketing the ground-state energy.

More broadly, our results indicate that Anderson-type lower bounds are a useful tool for the study of frustrated quantum magnets, particularly when combined with symmetry analysis and a judicious choice of clusters.
The approach developed here may be extended to larger motifs with computationally manageable Hilbert spaces, anisotropic interactions, and lattice geometries beyond the pyrochlore lattice.

\acknowledgments
We thank Sylvain Capponi, Yasir Iqbal, Masataka Kawano, Ilya Kull, and Norbert Schuch for valuable discussions. 
This work was supported by the Hungarian NKFIH OTKA Grant No. K 142652, by the Campus France scholarship “Séjour scientifique de haut niveau (SSHN)” No. FEH-2025-38 of the French Embassy, by the Intergovernmental Scholarship No. AK 2025-26/199932 funded by the Ministry of Culture and Innovation of Hungary.

\appendix

\section{Energies of the spin-1/2 $J_1-J_2$ Heisenberg model}

 In Table~\ref{tab:j3_0_spectrum} we give the analytical expressions for the levels minimizing the energy as we scan the free parameter for different values of $J_2$.

\begin{table*}[tb]
	\centering
	\caption{Analytic energies for the different $S_\text{tot}$ subspaces for the nearest and next nearest Heisenberg model Eq.~(\ref{eq:pyroham_J1J2J3}), where $J_{3b}=0$.  Indices refer to the regions in the energy diagram shown in Fig.~\ref{fig:2_nn_states}. For energies with a $\pm$, the minus sign corresponds to the lower energy. \label{tab:j3_0_spectrum}}
\begin{ruledtabular}
	\begin{tabular}{cccc}
		$S_\text{tot}$ & Irrep & Index & Energy \\ \hline
		$1/2$ & $A_{1g}$ & 1 & $-J_2-J' \pm \frac{1}{4}\sqrt{20J_2^2+9(1-2J')^2-20J_2(1-2J')}$ \\
		$1/2$ & $A_{2g}$ & 2 & $-\frac{3}{4}+\frac{3J_2}{2}+\frac{3J'}{2}$ \\
		$1/2$ & $E_{g}$ & 3 & $-\frac{3}{8}-\frac{J_2}{4}-\frac{J'}{4}\pm\frac{1}{8}\sqrt{68J_2^2+9(1-2J')^2-20J_2(1-2J')}$ \\
		$1/2$ & $A_{1u}$ & 4 & $-\frac{3}{4}+\frac{J_2}{2}+\frac{J'}{2}$ \\
		$1/2$ & $A_{2u}$ & 5 & $-J_2\pm\frac{1}{4}\sqrt{52J_2^2+9(1-2J')^2-36(1-2J')}$ \\
		$1/2$ & $E_u$ & -- & $-J_2-J'$ \\
		$1/2$ & $E_u$ & -- & $-\frac{3}{4}-\frac{J_2}{2}+\frac{3J'}{2}$ \\
		$3/2$ & $A_{1g}$ & 6 & $-J_2+\frac{J'}{2}\pm\frac{1}{4}\sqrt{20J_2^2+9(1-2J')^2-20J_2(1-2J')}$ \\
		$3/2$ & $E_g$ & 7 & $-\frac{3J'}{2}$ \\
		$3/2$ & $E_g$ & 8 & $-\frac{3}{8}-\frac{J_2}{4}+\frac{5J'}{4}\pm\frac{1}{8}\sqrt{68J_2^2+9(1-2J')^2-20J_2(1-2J')}$ \\
		$3/2$ & $A_{1u}$ & 9 & $-\frac{3}{4}+\frac{J_2}{2}+2J'$ \\
		$3/2$ & $A_{2u}$ & 10 & $\frac{3}{4}-\frac{J_2}{2}-3J'$ \\
		$3/2$ & $E_u$ & -- & $-J_2+\frac{J'}{2}$ \\
		$3/2$ & $E_u$ & -- & $J_2-\frac{3J'}{2}$ \\
		$5/2$ & $A_{1g}$ & 11 & $\frac{3}{4}+\frac{3J_2}{2}-\frac{7J'}{2}$ \\
		$5/2$ & $E_g$ & -- & $J'$ \\
		$5/2$ & $A_{2u}$ & -- & $\frac{3}{4}-\frac{J_2}{2}-\frac{J'}{2}$ \\
		$5/2$ & $E_u$ & -- & $J_2+J'$ \\
		$7/2$ & $A_{1g}$ & 12 & $\frac{3}{4}+\frac{3J_2}{2}$ 
	\end{tabular}
\end{ruledtabular}
\end{table*}

\section{Energies of the spin-1/2 $J_1-J_{23}$ Heisenberg model}

 In Table~\ref{tab:lower_Bound_J23} we give the analytical expressions for the levels minimizing the energy as we scan the free parameter for different values of the special $J_2 = J_{3b}\equiv J_{23}$ case.

\begin{table}[!tb]
\caption{
Rigorous lower bounds $E_{\mathrm{LB}}$ on the ground-state energy of the $J_1$--$J_2$--$J_{3b}$ Heisenberg model on the pyrochlore lattice in the symmetric case $J_2=J_{3b}\equiv J_{23}$, obtained from the seven-site hourglass construction.
The parameter space is divided into intervals of $J_{23}/J_1$ separated by level crossings of the hourglass Hamiltonian.
For each interval, we list the optimal value of the free parameter $J'$, the corresponding lower bound $e_\mathrm{LB}$, and the spin quantum numbers
$\Sigma=(S_{234},S_{567},S_{234567},S_{\mathrm{tot}})$
of the hourglass ground state, together with their multiplicities. The star * in front of the multiplicities denotes whether the spin multiplet satisfies the Lieb-Mattis theorem. The FM denotes the ferromagnetic state, whose energy is independent of the $J'$ free parameter -- the $J'$ can take values $-6J_{23} \leq J'\leq 0$, bounded by crossings with $S_{\mathrm{tot}}=5/2$ levels.
\label{tab:lower_Bound_J23}}
\begin{ruledtabular}
\begin{tabular}{lccr}
$J' $ & $ e_\mathrm{LB} $ &
$ \Sigma $ & mult.\\ 
\hline
\multicolumn{2}{c}{$J_{23} =  0$, $J_1>0$} \\[0.4em]
%
\multirow{3}{*}{$ \frac{3}{8} J_1 +\frac{1}{4} J_{23}$} &
\multirow{3}{*}{$ -\frac{9}{16} J_1  + \frac{3}{8} J_{23} $} &
$ ( 1/2,1/2,1,1/2) $ & $ *\,4 $ \\
& & 
$ ( 1/2,3/2,2,3/2) $ & $ *\,2 $ \\
& &
$ (3/2,1/2,2,3/2)$ & $ *\,2 $ \\
\multicolumn{2}{c}{$J_{23} =  \frac{1}{10} J_1 >0  $} \\[0.4em]
%
\multirow{4}{*}{$ \frac{1}{2} J_1 -J_{23} $} &
\multirow{4}{*}{$ -\frac{1}{2} J_1-\frac{1}{4} J_{23} $} &
$ (1/2,1/2,1,1/2)$ & $ *\,4 $\\
& & 
$ (1/2,3/2,1,1/2)$ & $ 2 $ \\
& & 
$ (3/2,1/2,1,1/2)$ & $ 2 $ \\
& &
$ (3/2,3/2,1,1/2)$ & $ 1 $ \\
\multicolumn{2}{c}{$J_{23} = \frac{1}{4} J_1 >0 $} \\[0.4em]
%
\multirow{2}{*}{$ \frac{1}{2} J_1 -J_{23} $} &
\multirow{2}{*}{$ -\frac{9}{4} J_{23} $} &
$ (1/2,1/2,0,1/2)$ & $ 4 $ \\
& & 
$ (3/2,3/2,0,1/2)$ & $ 1 $ \\
\multicolumn{2}{c}{$J_{23} =  -\frac{1}{2} J_1$, $J_1<0$} \\[0.4em]
%
\multirow{4}{*}{$ \frac{1}{2} J_1 -J_{23} $} &
\multirow{4}{*}{$ \frac{1}{4} J_1 -\frac{7}{4} J_{23}$} &
$ (1/2,1/2,1,3/2)$ & $ 4 $ \\
& & 
$ (1/2,3/2,1,3/2)$ & $ 2 $ \\
& & 
$ (3/2,1/2,1,3/2)$ & $ 2 $ \\
& & 
$ (3/2,3/2,1,3/2)$ & $ 1 $ \\
\multicolumn{2}{c}{$J_{23} =  -\frac{1}{6} J_1$, $J_1<0$} \\[0.4em]
%
\multirow{3}{*}{$ \frac{1}{2} J_1 -J_{23} $} &
\multirow{3}{*}{$ \frac{1}{2} J_1 -\frac{1}{4} J_{23} $} &
$ (1/2,3/2,2,5/2)$ & $ 2 $ \\
& & 
$ (3/2,1/2,2,5/2)$ & $ 2 $ \\
& & 
$ (3/2,3/2,2,5/2)$ & $ 1 $ \\
\multicolumn{2}{c}{$J_{23} =  -\frac{1}{10} J_1 $, $J_1<0$} \\[0.4em]
%
\multirow{3}{*}{$ -6J_{23} \leq J'\leq 0 $} &
\multirow{3}{*}{$ \frac{3}{4} J_1 +\frac{9}{4} J_{23} $} &
$ (3/2,3/2,2,5/2)$ & $ *\,1 $ \\
& & 
$ (3/2,3/2,3,7/2)$ & FM$\, 1 $ \\
& &
$ (3/2,3/2,3,5/2)$ & $ 1 $ \\
\multicolumn{2}{c}{$J_{23} =  -\frac{3}{4} J_1 $, $J_1>0$} \\[0.4em]
%
\multirow{2}{*}{$ \frac{3}{8} J_1 +\frac{1}{2} J_{23} $} &
\multirow{2}{*}{$ -\frac{9}{16} J_1 + \frac{1}{2} J_{23} $} &
$ (1/2,1/2,1,1/2)$ & $ *\,4 $ \\
& & 
$ (3/2,3/2,3,5/2)$ & $ *\,1 $ \\
\multicolumn{2}{c}{$J_{23} =  0$, $J_1>0$} \\[0.4em]
\end{tabular}
\end{ruledtabular}
\end{table}

\section{4-site ring exchanges}
\label{sec:foursite_appendix}

\begin{table}[tb]
	\caption{Analytic energies in different $S_\text{tot}$ subspaces for the bilinear-quadrilinear model, Eq.~(\ref{eq:ringexchangewithP}).  Indices label the regions in the energy diagram shown in Fig.~\ref{fig:4ex_energies_states}. For energies having a $\pm$ sign, the minus sign corresponds to the lower-energy branch.
	\label{tab:4site_exchange_energies_explicit}}
	\begin{ruledtabular}
	\begin{tabular}{cccc}
		$S_\text{tot}$ & Irrep & Index & Energy \\ \hline
		1/2 & $A_{1g}$ & -- & $\frac{3}{4}-\frac{7K}{32}-5\tilde{J}$ \\
		1/2 & $E_g,\;E_u$ & 1 & $-\frac{7K}{32}-2\tilde{J}$ \\
		1/2 & $A_{1g},\;A_{1u},\;E_g$ & 2 & $-\frac{3}{4}+\frac{11K}{32}+\tilde{J}$ \\
		1/2 & $A_{2u}$ & -- & $\frac{3}{4}-\frac{3K}{32}-3\tilde{J}$ \\
		1/2 & $A_{2u},\;A_{2g},\;E_u$ & 3 & $-\frac{3}{4}+\frac{3K}{32}+3\tilde{J}$ \\
		3/2 & $A_{2u}$ & -- & $\frac{3}{4}-\frac{9K}{32}-6\tilde{J}$ \\
		3/2 & $A_{1g}$ & -- & $\frac{3}{4}-\frac{K}{32}-2\tilde{J}$ \\
		3/2 & $A_{1g},\;A_{1u},\;E_g$ & 4 & $-\frac{3}{4}-\frac{K}{32}+4\tilde{J}$ \\
		3/2 & $E_g,\;E_u$ & 5 & 
$\begin{array}{l}
\frac{K}{32}-\tilde{J} \\
\quad\pm \frac{1}{8\sqrt{2}}\sqrt{5K^2+40K\tilde{J}+512 \tilde{J}^2}
\end{array}$\\
		5/2 & $A_{2u}$ & -- & $\frac{3}{4}+\frac{K}{32}-\tilde{J}$ \\
		5/2 & $A_{1g}$ & 6 & $\frac{3}{4}-\frac{11K}{32}-7\tilde{J}$ \\
		5/2 & $E_g,\;E_u$ & -- & $\frac{K}{32}+2\tilde{J}$ \\
		7/2 & $A_{1g}$ & -- & $\frac{3}{4}+\frac{3K}{32}$ 
	\end{tabular}
  \end{ruledtabular}
\end{table}

For computational convenience, we rewrite the spin bilinears in terms of permutation operators,
\begin{equation}
	\mathcal{P}_{i,j}
	=
	2\,\mathbf{S}_i\cdot \mathbf{S}_j
	+
	\frac{1}{2},
\end{equation}
and introduce the shifted coupling
\begin{equation}
   \label{eq:Jtildedef}
	\tilde{J}
	=
	\frac{J^\prime}{2}
	-
	\frac{K}{16}.
\end{equation}
With these definitions, the hourglass sub-Hamiltonian
Eq.~(\ref{eq:motifham7_4exchange}) takes the form
\begin{multline}
\label{eq:ringexchangewithP}
	\mathcal{H}_0= 
	 -\frac{3J}{4}+\frac{3K}{32}
	 \\
	 +\tilde{J}\left(\mathcal{P}_{1,2}+\mathcal{P}_{1,3}+\mathcal{P}_{1,4}+\mathcal{P}_{1,5}+\mathcal{P}_{1,6}+\mathcal{P}_{1,7}\right)\ \\
	+ \left(\frac{J}{4} \!-\! \frac{K}{16} \!-\! \tilde{J}\right)\left(\mathcal{P}_{2,3}+\mathcal{P}_{2,4}+\mathcal{P}_{3,4}+\mathcal{P}_{5,6}+\mathcal{P}_{5,7}+\mathcal{P}_{6,7}\right)
	\\
	+\frac{K}{16}\left(\mathcal{P}_{1,2}\mathcal{P}_{3,4}+\mathcal{P}_{1,3}\mathcal{P}_{2,4}+\mathcal{P}_{1,4}\mathcal{P}_{2,3} 
	\right.\\ \left.
	+\mathcal{P}_{1,5}\mathcal{P}_{6,7}+\mathcal{P}_{1,6}\mathcal{P}_{5,7}+\mathcal{P}_{1,7}\mathcal{P}_{5,6}\right).
\end{multline}

Using this representation, we obtain closed-form expressions for all energy eigenvalues of the spin-$\tfrac12$ hourglass Hamiltonian with bilinear and quadrilinear couplings.
Table~\ref{tab:4site_exchange_energies_explicit} lists the analytic energies in the
$S_\mathrm{tot}=1/2$, $3/2$, $5/2$, and $7/2$ sectors, together with their corresponding
irreducible representations of the $D_{3d}$ point group.

For each entry, the total degeneracy is given by
$(2S_\mathrm{tot}+1)\times \dim(\Gamma)$.
Indices label the branches used in constructing the phase diagrams shown in
Fig.~\ref{fig:4ex_energies_states}.
For energies containing a $\pm$ sign, the lower branch corresponds to the minus sign.

\section{Crown cluster Hamiltonian}
\label{sec:appendix_crown}

In this appendix, we give the explicit form of the most general bilinear spin Hamiltonian defined on the 18-site crown cluster shown in Fig.~\ref{fig:crown}, together with the constraints imposed by embedding this motif into the pyrochlore lattice.

\subsection{Symmetry-inequivalent bond orbits}

The crown cluster admits 19 symmetry-inequivalent orbits of spin–spin bilinears under the $D_{3d}$ point-group symmetries. Each orbit $o_m$ consists of a set of equivalent bonds $(i,j)$ that share a common coupling constant $c_m$. Table~\ref{tab:crown_orbits} lists these orbits explicitly.

The most general bilinear Hamiltonian on the crown cluster can then be written as
\begin{equation}
	\mathcal{H}_0 = \sum_{m=1}^{19} \sum_{(i,j)\in o_m} c_m\, \mathbf{S}_i \cdot \mathbf{S}_j .
\end{equation}
When embedding the crown cluster into the pyrochlore lattice, the coupling constants are constrained by matching to the lattice Hamiltonian with nearest-, second-, and two symmetry-inequivalent third-neighbor exchanges.
We denote by $J_{3a}$ the third-neighbor coupling across hexagonal loops, and by $J_{3b}$ the third-neighbor coupling between sites belonging to corner-sharing tetrahedra.
The matching conditions then read
\begin{subequations}
\begin{align}
	J = J_1 &=2c_1+4c_{3}+4c_{12}+2c_{17} \,,\\
	J_2&=2 c_{2}+2c_{5}+c_{11}+c_{18}\,,\\
	J_{3a}&=2c_{4}+c_{19} \,,\\
	J_{3b}&=2c_{6}+4c_{13} \,,\\
	0&=c_7+c_{15} \,,\\
	0&=c_8=c_{9}=c_{10}=c_{14}=c_{16}\,.
\end{align}
\end{subequations}
The constraints set to zero correspond to longer-range exchange processes that are allowed by the symmetry of the crown cluster but are absent in the original lattice Hamiltonian (\ref{eq:heisenberg_model}). With these constraints imposed, the crown Hamiltonian takes the explicit form
\begin{widetext}
\begin{multline}
	\mathcal{H}_0= c_1(\mathbf{S}_{1}\cdot\mathbf{S}_{2}+\mathbf{S}_{3}\cdot\mathbf{S}_{10}+\mathbf{S}_{6}\cdot\mathbf{S}_{12}+\mathbf{S}_{7}\cdot\mathbf{S}_{13}+\mathbf{S}_{9}\cdot\mathbf{S}_{16}+\mathbf{S}_{17}\cdot\mathbf{S}_{18})\\
	+c_{2}(\mathbf{S}_{1}\cdot\mathbf{S}_{3}+\mathbf{S}_{1}\cdot\mathbf{S}_{6}+\mathbf{S}_{2}\cdot\mathbf{S}_{10}+\mathbf{S}_{2}\cdot\mathbf{S}_{12}+\mathbf{S}_{3}\cdot\mathbf{S}_{7}+\mathbf{S}_{6}\cdot\mathbf{S}_{9} 	
+\mathbf{S}_{7}\cdot\mathbf{S}_{17}+\mathbf{S}_{9}\cdot\mathbf{S}_{17}+\mathbf{S}_{10}\cdot\mathbf{S}_{13}+\mathbf{S}_{12}\cdot\mathbf{S}_{16}+\mathbf{S}_{13}\cdot\mathbf{S}_{18}+\mathbf{S}_{16}\cdot\mathbf{S}_{18})
\\	
+c_{3}(\mathbf{S}_{1}\cdot\mathbf{S}_{4}+\mathbf{S}_{1}\cdot\mathbf{S}_{5}+\mathbf{S}_{4}\cdot\mathbf{S}_{10}+\mathbf{S}_{5}\cdot\mathbf{S}_{12}+\mathbf{S}_{7}\cdot\mathbf{S}_{8}+\mathbf{S}_{7}\cdot\mathbf{S}_{14}	
+\mathbf{S}_{8}\cdot\mathbf{S}_{10}+\mathbf{S}_{9}\cdot\mathbf{S}_{11}+\mathbf{S}_{9}\cdot\mathbf{S}_{15}+\mathbf{S}_{11}\cdot\mathbf{S}_{12}+\mathbf{S}_{14}\cdot\mathbf{S}_{18}+\mathbf{S}_{15}\cdot\mathbf{S}_{18})
\\	
+c_{4}(\mathbf{S}_{1}\cdot\mathbf{S}_{7}+\mathbf{S}_{1}\cdot\mathbf{S}_{9}+\mathbf{S}_{7}\cdot\mathbf{S}_{9}+\mathbf{S}_{10}\cdot\mathbf{S}_{12}+\mathbf{S}_{10}\cdot\mathbf{S}_{18}+\mathbf{S}_{12}\cdot\mathbf{S}_{18})
\\	
+c_{5}(\mathbf{S}_{1}\cdot\mathbf{S}_{8}+\mathbf{S}_{1}\cdot\mathbf{S}_{11}+\mathbf{S}_{4}\cdot\mathbf{S}_{7}+\mathbf{S}_{4}\cdot\mathbf{S}_{12}+\mathbf{S}_{5}\cdot\mathbf{S}_{9}+\mathbf{S}_{5}\cdot\mathbf{S}_{10}
+\mathbf{S}_{7}\cdot\mathbf{S}_{15}+\mathbf{S}_{8}\cdot\mathbf{S}_{18}+\mathbf{S}_{9}\cdot\mathbf{S}_{14}+\mathbf{S}_{10}\cdot\mathbf{S}_{14}+\mathbf{S}_{11}\cdot\mathbf{S}_{18}+\mathbf{S}_{12}\cdot\mathbf{S}_{15})
\\	
+c_{6}(\mathbf{S}_{1}\cdot\mathbf{S}_{10}+\mathbf{S}_{1}\cdot\mathbf{S}_{12}+\mathbf{S}_{7}\cdot\mathbf{S}_{10}+\mathbf{S}_{7}\cdot\mathbf{S}_{18}+\mathbf{S}_{9}\cdot\mathbf{S}_{12}+\mathbf{S}_{9}\cdot\mathbf{S}_{18})
\\	
+c_{7}(\mathbf{S}_{1}\cdot\mathbf{S}_{13}+\mathbf{S}_{1}\cdot\mathbf{S}_{16}+\mathbf{S}_{2}\cdot\mathbf{S}_{7}+\mathbf{S}_{2}\cdot\mathbf{S}_{9}+\mathbf{S}_{3}\cdot\mathbf{S}_{12}+\mathbf{S}_{3}\cdot\mathbf{S}_{18}
+\mathbf{S}_{6}\cdot\mathbf{S}_{10}+\mathbf{S}_{6}\cdot\mathbf{S}_{18}+\mathbf{S}_{7}\cdot\mathbf{S}_{16}+\mathbf{S}_{9}\cdot\mathbf{S}_{13}+\mathbf{S}_{10}\cdot\mathbf{S}_{17}+\mathbf{S}_{12}\cdot\mathbf{S}_{17})
\\	
+c_{11}(\mathbf{S}_{2}\cdot\mathbf{S}_{3}+\mathbf{S}_{2}\cdot\mathbf{S}_{6}+\mathbf{S}_{3}\cdot\mathbf{S}_{13}+\mathbf{S}_{6}\cdot\mathbf{S}_{16}+\mathbf{S}_{13}\cdot\mathbf{S}_{17}+\mathbf{S}_{16}\cdot\mathbf{S}_{17})
\\	
+c_{12}(\mathbf{S}_{2}\cdot\mathbf{S}_{4}+\mathbf{S}_{2}\cdot\mathbf{S}_{5}+\mathbf{S}_{3}\cdot\mathbf{S}_{4}+\mathbf{S}_{3}\cdot\mathbf{S}_{8}+\mathbf{S}_{5}\cdot\mathbf{S}_{6}+\mathbf{S}_{6}\cdot\mathbf{S}_{11}
+\mathbf{S}_{8}\cdot\mathbf{S}_{13}+\mathbf{S}_{11}\cdot\mathbf{S}_{16}+\mathbf{S}_{13}\cdot\mathbf{S}_{14}+\mathbf{S}_{14}\cdot\mathbf{S}_{17}+\mathbf{S}_{15}\cdot\mathbf{S}_{16}+\mathbf{S}_{15}\cdot\mathbf{S}_{17})
\\	-\frac{c_6}{2}(\mathbf{S}_{2}\cdot\mathbf{S}_{8}+\mathbf{S}_{2}\cdot\mathbf{S}_{11}+\mathbf{S}_{3}\cdot\mathbf{S}_{5}+\mathbf{S}_{3}\cdot\mathbf{S}_{14}+\mathbf{S}_{4}\cdot\mathbf{S}_{6}+\mathbf{S}_{4}\cdot\mathbf{S}_{13}
	+\mathbf{S}_{5}\cdot\mathbf{S}_{16}+\mathbf{S}_{6}\cdot\mathbf{S}_{15}+\mathbf{S}_{8}\cdot\mathbf{S}_{17}+\mathbf{S}_{11}\cdot\mathbf{S}_{17}+\mathbf{S}_{13}\cdot\mathbf{S}_{15}+\mathbf{S}_{14}\cdot\mathbf{S}_{16})
\\	
-c_7(\mathbf{S}_{2}\cdot\mathbf{S}_{14}+\mathbf{S}_{2}\cdot\mathbf{S}_{15}+\mathbf{S}_{3}\cdot\mathbf{S}_{11}+\mathbf{S}_{3}\cdot\mathbf{S}_{15}+\mathbf{S}_{4}\cdot\mathbf{S}_{16}+\mathbf{S}_{4}\cdot\mathbf{S}_{17}
	+\mathbf{S}_{5}\cdot\mathbf{S}_{13}+\mathbf{S}_{5}\cdot\mathbf{S}_{17}+\mathbf{S}_{6}\cdot\mathbf{S}_{8}+\mathbf{S}_{6}\cdot\mathbf{S}_{14}+\mathbf{S}_{8}\cdot\mathbf{S}_{16}+\mathbf{S}_{11}\cdot\mathbf{S}_{13})
\\	+\left(\frac{J}{2}-c_1-2c_3-2c_{12}\right)(\mathbf{S}_{4}\cdot\mathbf{S}_{5}+\mathbf{S}_{4}\cdot\mathbf{S}_{8}+\mathbf{S}_{5}\cdot\mathbf{S}_{11}+\mathbf{S}_{8}\cdot\mathbf{S}_{14}
	+\mathbf{S}_{11}\cdot\mathbf{S}_{15}+\mathbf{S}_{14}\cdot\mathbf{S}_{15})
\\ -2\left(c_2+c_5+\frac{c_{11}}{2}\right)(\mathbf{S}_{4}\cdot\mathbf{S}_{11}+\mathbf{S}_{4}\cdot\mathbf{S}_{14}+\mathbf{S}_{5}\cdot\mathbf{S}_{8}
	+\mathbf{S}_{5}\cdot\mathbf{S}_{15}+\mathbf{S}_{8}\cdot\mathbf{S}_{15}+\mathbf{S}_{11}\cdot\mathbf{S}_{14})
	\\
-2c_4(\mathbf{S}_{4}\cdot\mathbf{S}_{15}+\mathbf{S}_{5}\cdot\mathbf{S}_{14}+\mathbf{S}_{8}\cdot\mathbf{S}_{11}).
	\label{eq:crown_hamiltonian}
\end{multline}
\end{widetext}

\begin{table}[tb]
  \caption{Symmetry-inequivalent bond orbits of the 18-site crown cluster. Each orbit $o_m$ consists of symmetry-related bonds sharing a common coupling constant $c_m$.}
  \label{tab:crown_orbits}
  \begin{ruledtabular}
  \begin{tabular}{ccl}
	Orbit    & $\mathrm{S}_i \mathrm{S}_j$                                                                    \\ 
	\hline
	$o_1$    & (1,2), (3,10), (6,12), (7,13), (9,16), (17,18)                                                 \\
\multirow{2}{*}{$o_2$} & (1,3), (1,6), (2,10), (2,12), (3,7), (6,9), (7,17), \\
& (9,17), (10,13), (12,16), (13,18), (16,18) \\
\multirow{2}{*}{$o_3$} & (1,4), (1,5), (4,10), (5,12), (7,8), (7,14), (8,10), \\
& (9,11), (9,15), (11,12), (14,18), (15,18) \\
	$o_4$    & (1,7), (1,9), (7,9), (10,12), (10,18), (12,18)                                                 \\
\multirow{2}{*}{$o_5$} & (1,8), (1,11), (4,7), (4,12), (5,9), (5,10), (7,15), \\
& (8,18), (9,14), (10,14), (11,18), (12,15) \\
	$o_6$    & (1,10), (1,12), (7,10), (7,18), (9,12), (9,18)                                                 \\
\multirow{2}{*}{$o_7$} & (1,13), (1,16), (2,7), (2,9), (3,12), (3,18), (6,10), \\
& (6,18), (7,16), (9,13), (10,17), (12,17) \\
\multirow{2}{*}{$o_8$} & (1,14), (1,15), (4,9), (4,18), (5,7), (5,18), (7,11), \\
& (8,9), (8,12), (10,11), (10,15), (12,14) \\
	$o_9$    & (1,17), (2,18), (3,9), (6,7), (10,16), (12,13)                                                 \\
	$o_{10}$ & (1,18), (7,12), (9,10)                                                                         \\
	$o_{11}$ & (2,3), (2,6), (3,13), (6,16), (13,17), (16,17)                                                 \\
\multirow{2}{*}{$o_{12}$} & (2,4), (2,5), (3,4), (3,8), (5,6), (6,11), (8,13), \\
& (11,16), (13,14), (14,17), (15,16), (15,17) \\
\multirow{2}{*}{$o_{13}$} & (2,8), (2,11), (3,5), (3,14), (4,6), (4,13), (5,16), \\
& (6,15), (8,17), (11,17), (13,15), (14,16) \\
	$o_{14}$ & (2,13), (2,16), (3,6), (3,17), (6,17), (13,16)                                                 \\
\multirow{2}{*}{$o_{15}$} & (2,14), (2,15), (3,11), (3,15), (4,16), (4,17), (5,13), \\
& (5,17), (6,8), (6,14), (8,16), (11,13) \\
	$o_{16}$ & (2,17), (3,16), (6,13)                                                                         \\
	$o_{17}$ & (4,5), (4,8), (5,11), (8,14), (11,15), (14,15)                                                 \\
	$o_{18}$ & (4,11), (4,14), (5,8), (5,15), (8,15), (11,14)                                                 \\
	$o_{19}$ & (4,15), (5,14), (8,11) 
	\end{tabular}
  \end{ruledtabular}
\end{table}

Optimizing the ground-state energy of the crown Hamiltonian, we find the lower bound
\begin{equation}
e_\mathrm{LB} = -0.549832(8),
\end{equation}
for the coupling constants
\begin{subequations}
\label{eq:crown_cs}
\begin{align}
	c_1=&0.037416(7)  \,,\\
	c_{2}=&0.001151(3) \,,\\
	c_{3}=&0.083729(9) \,,\\
	c_{4}=&0.0000415(5)  \,,\\
	c_{5}=&-0.000546(9)  \,,\\
	c_{6}=&-0.000210(9) \,,\\
	c_{7}=&-0.000539(3)  \,,\\
	c_{11}=&0.001764(2)  \,,\\
	c_{12}=&0.080243(4) \,.
	\label{eq:pars}
\end{align}
\end{subequations}

Under the additional constraints discussed in the main text, the squared total spin of each outer spin pair, 
$\tilde{\mathbf S}_{ij}^2=(\mathbf S_i+\mathbf S_j)^2$,
commutes with $\mathcal{H}_0$.
This allows each outer pair to be replaced by an effective spin $\tilde S_{ij}=0,1,\ldots,2S$,
leading to the reduced Hamiltonian
\begin{multline}
  \label{eq:crown_red}
	\mathcal{H}_0^{(\mathrm{red.})} = 
	\left(\frac{J}{2}-c_1-4c_3\right) (\mathbf{S}_8\cdot\mathbf{S}_4+\mathbf{S}_4\cdot\mathbf{S}_5
	\\ +\mathbf{S}_5\cdot\mathbf{S}_{11}+\mathbf{S}_{11}\cdot\mathbf{S}_{15}+\mathbf{S}_{15}\cdot\mathbf{S}_{14}+\mathbf{S}_{14}\cdot\mathbf{S}_8) 
	\\
	+c_3(
	\mathbf{\tilde S}_{7,13}\cdot\mathbf{S}_8+\mathbf{\tilde S}_{3,10}\cdot\mathbf{S}_8+\mathbf{\tilde S}_{3,10}\cdot\mathbf{S}_4+\mathbf{\tilde S}_{1,2}\cdot\mathbf{S}_4
	\\
	+\mathbf{\tilde S}_{1,2}\cdot\mathbf{S}_5+\mathbf{\tilde S}_{6,12}\cdot\mathbf{S}_5+\mathbf{\tilde S}_{6,12}\cdot\mathbf{S}_{11}+\mathbf{\tilde S}_{9,16}\cdot\mathbf{S}_{11}
	\\
	+\mathbf{\tilde S}_{9,16}\cdot\mathbf{S}_{15}+\mathbf{\tilde S}_{17,18}\cdot\mathbf{S}_{15}+\mathbf{\tilde S}_{17,18}\cdot\mathbf{S}_{14}+\mathbf{\tilde S}_{7,13}\cdot\mathbf{S}_{14})
	\\
	+c_1(E_{1,2}+E_{6,12}+E_{9,16}+E_{17,18}+E_{7,13}+E_{3,10}),
\end{multline}
where
\begin{equation}
	E_{i,j}=\frac{1}{2}\tilde{S}_{i,j}(\tilde{S}_{i,j}+1)-S(S+1)
\end{equation}
takes into account the energy of the fused outer spin pairs.

This reduction leads to a dramatic decrease in Hilbert-space dimension.
Imposing the conservation of the outer-pair spins $\tilde S_{i,j}$ decomposes the Hilbert space into sectors labeled by the spin configuration of the outer pairs.

For $S=\tfrac12$, each outer pair can form either a singlet ($\tilde S_{i,j}=0$) or a triplet ($\tilde S_{i,j}=1$).
If all six outer pairs are singlets, the $S_\mathrm{tot}^z=0$ sector has dimension only $20$.
With one triplet and five singlets, the dimension increases to $50$, and it grows monotonically as additional outer pairs are promoted to triplets.
The largest sector occurs when all six outer pairs are triplets, yielding a dimension of $7780$.
In all cases, this represents a drastic reduction compared to the full $S_\mathrm{tot}^z=0$ Hilbert space of dimension $48\,620$ for the unrestricted 18-site cluster.
\subsection{Sub-Hamiltonian for the chiral Hamiltonian}
In case of the uniform chiral Eq.~(\ref{eq:pyroham_chiral_uniform}) and alternating chiral Eq.~(\ref{eq:pyroham_chiral_alternating}) Hamiltonians, for the chiral interactions on the tetrahedra, there are 4 symmetry-inequivalent orbits both for the $C_{3i}$ and $D_3$ symmetries. Each orbit $o_m$ consists of a set of equivalent oriented triangles $(i,j,k)$ that share a common coupling constant $k_m$. Table~\ref{tab:crown_orbits_chiral} lists these orbits explicitly.

The Hamiltonian with these interactions can be written as
\begin{equation}
	\mathcal{H}_{\chi_0} = \sum_{m=1}^{4} \sum_{(i,j,k)\in o_m} k_m\, \chi_{i,j,k} .
\end{equation}
When embedding the crown cluster into the pyrochlore lattice, the coupling constants are constrained by matching to the lattice Hamiltonian, Eq.~(\ref{eq:pyroham_chiral_uniform}) for the uniform case with the $D_{3}$ symmetry and Eq.~(\ref{eq:pyroham_chiral_alternating}) for the alternating case with  $C_{3i}$ symmetry. The embedding constraint in both cases is
\begin{equation}
	J_{\chi}=-12(k_1+k_2+k_3+k_4).
\end{equation}

\begin{table}[tb]
	\caption{Symmetry-inequivalent chirality  orbits of the 18-site crown cluster in case of the $D_3$ and $C_{3i}$ groups. Each orbit $o_m$ consists of symmetry-related triangles sharing a common coupling constant $k_m$.}
	\label{tab:crown_orbits_chiral}
	\begin{tabular}{cc}
		\hline\hline
		$D_{3}$ Orbit  & $\chi_{i,j,k}$                                                                            \\ \hline
		$o_1$          & \begin{tabular}[c]{@{}c@{}}(1,2,4), (3,4,10), (6,11,12), \\ (7,13,14), (9,16,11), (14,18,17)\end{tabular} \\
		$o_2$          & \begin{tabular}[c]{@{}c@{}}(1,4,5), (4,8,10), (5,12,11), \\ (7,14,8), (9,11,15), (14,15,18)\end{tabular}  \\
		$o_3$          & \begin{tabular}[c]{@{}c@{}}(1,5,2), (3,10,8), (5,6,12),\\ (7,8,13), (9,15,16), (15,17,18)\end{tabular}    \\
		$o_4$          & \begin{tabular}[c]{@{}c@{}}(2,5,4), (3,8,4), (5,11,6),\\ (8,14,13), (11,16,15), (14,17,15)\end{tabular}   \\ \hline\hline
		$C_{3i}$ Orbit & $\chi_{i,j,k}$                                                                            \\ \hline
		$o_1$          & \begin{tabular}[c]{@{}c@{}}(1,2,4), (3,8,10), (5,12,6), \\ (7,13,14), (9,16,11), (15,18,17)\end{tabular}  \\
		$o_2$          & \begin{tabular}[c]{@{}c@{}}(1,4,5), (4,10,8), (5,11,12), \\ (7,14,8), (9,11,15), (14,18,15)\end{tabular}  \\
		$o_3$          & \begin{tabular}[c]{@{}c@{}}(1,5,2), (3,10,4), (6,12,11), \\ (7,8,13), (9,15,16), (14,17,18)\end{tabular}  \\
		$o_4$          & \begin{tabular}[c]{@{}c@{}}(2,5,4), (3,4,8), (5,6,11), \\ (8,14,13), (11,16,15), (14,15,17)\end{tabular}  \\ \hline\hline
	\end{tabular}
\end{table}

For the uniform case, this yields the Hamiltonian
\begin{align}
	\mathcal{H}_{\chi_0}^{\mathrm{(uni)}}=& \left(\frac{J_{\chi}}{3}+4k_1+4k_2+4k_3\right)(\chi_{2,4,5}+\chi_{3,4,8}\nonumber\\
	&+\chi_{5,6,11}+\chi_{8,13,14}+\chi_{11,15,16}+\chi_{14,15,17}) \nonumber\\
	&+4 k_1 (\chi_{1,2,4}+\chi_{3,4,10}+\chi_{6,11,12}\nonumber\\
	&+\chi_{7,13,14}+\chi_{9,16,11}+\chi_{14,18,17})\nonumber\\
	&+4k_2(\chi_{1,4,5}+\chi_{4,8,10}+\chi_{5,12,11}\nonumber\\
	&+\chi_{7,14,8}+\chi_{9,11,15}+\chi_{14,15,18})\nonumber\\
	&+4k_3(\chi_{1,5,2}+\chi_{3,10,8}+\chi_{5,6,12}\nonumber\\
	&+\chi_{7,8,13}+\chi_{9,15,16}+\chi_{15,17,18}) 
	\label{eq:crown_hamiltonian_chiral}
\end{align}
and for the alternating case
\begin{align}
	\mathcal{H}_{\chi_0}^{\mathrm{(alt)}}=& \left(\frac{J_{\chi}}{3}+4k_1+4k_2+4k_3\right)(\chi_{2,4,5}+\chi_{3,8,4}\nonumber\\
	&+\chi_{5,11,6}+\chi_{8,13,14}+\chi_{11,15,16}+\chi_{14,17,15}) \nonumber\\
	&+4 k_1 (\chi_{1,2,4}+\chi_{3,8,10}+\chi_{5,12,6}\nonumber\\
	&+\chi_{7,13,14}+\chi_{9,16,11}+\chi_{15,18,17})\nonumber\\
	&+4k_2(\chi_{1,4,5}+\chi_{4,10,8}+\chi_{5,11,12}\nonumber\\
	&+\chi_{7,14,8}+\chi_{9,11,15}+\chi_{14,18,15})\nonumber\\
	&+4k_3(\chi_{1,5,2}+\chi_{3,10,4}+\chi_{6,12,11}\nonumber\\
	&+\chi_{7,8,13}+\chi_{9,15,16}+\chi_{14,17,18}) .
	\label{eq:crown_hamiltonian_chiral_alt}
\end{align}

Optimizing the ground-state energy of the uniform Hamiltonian, we found the lower bound
\begin{equation}
	e_\mathrm{LB} = -0.481197(6),
\end{equation}
with the coupling constants
\begin{subequations}
	\label{eq:crown_cs_chiral}
	\begin{align}
		k_1=&-0.005999(1)\,,\\
		k_2=&-0.034479(2)\,,\\
		k_3=&-0.005282(2)\,,\\
		c_1=&-0.023143(8)  \,,\\
		c_{2}=&-0.000879(2) \,,\\
		c_{3}=&-0.000829(2)\,,\\
		c_{4}=&-0.001489(1)  \,,\\
		c_{5}=&0.003700(8)  \,,\\
		c_{6}=&0.001466(8) \,,\\
		c_{7}=&0.000500(8)  \,,\\
		c_{11}=&-0.001543(9)  \,,\\
		c_{12}=&-0.001365(7) \,.
		\label{eq:pars_chi}
	\end{align}
\end{subequations}

For the alternating Hamiltonian, we got
\begin{equation}
	e_\mathrm{LB} =-0.481441(7),
\end{equation}
with the parameters
\begin{subequations}
	\label{eq:crown_cs_chiral_alt}
	\begin{align}
		k_1=&-0.006316(2)\,,\\
		k_2=&-0.034659(5)\,,\\
		k_3=&-0.006398(1)\,,\\
		c_1=&-0.024213(6)  \,,\\
		c_{2}=&0.002348(7) \,,\\
		c_{3}=&-0.001320(1)\,,\\
		c_{4}=&-0.001679(3)  \,,\\
		c_{5}=&0.000529(7)  \,,\\
		c_{6}=&-0.001019(1) \,,\\
		c_{7}=&-0.000206(2)  \,,\\
		c_{11}=&0.000595(4)  \,,\\
		c_{12}=&0.000745(5) \,.
		\label{eq:pars_chi_alt}
	\end{align}
\end{subequations}

\bibliography{bib_pyrochlore}

\begin{thebibliography}{32}%
\makeatletter
\providecommand \@ifxundefined [1]{%
 \@ifx{#1\undefined}
}%
\providecommand \@ifnum [1]{%
 \ifnum #1\expandafter \@firstoftwo
 \else \expandafter \@secondoftwo
 \fi
}%
\providecommand \@ifx [1]{%
 \ifx #1\expandafter \@firstoftwo
 \else \expandafter \@secondoftwo
 \fi
}%
\providecommand \natexlab [1]{#1}%
\providecommand \enquote  [1]{``#1''}%
\providecommand \bibnamefont  [1]{#1}%
\providecommand \bibfnamefont [1]{#1}%
\providecommand \citenamefont [1]{#1}%
\providecommand \href@noop [0]{\@secondoftwo}%
\providecommand \href [0]{\begingroup \@sanitize@url \@href}%
\providecommand \@href[1]{\@@startlink{#1}\@@href}%
\providecommand \@@href[1]{\endgroup#1\@@endlink}%
\providecommand \@sanitize@url [0]{\catcode `\\12\catcode `\$12\catcode
  `\&12\catcode `\#12\catcode `\^12\catcode `\_12\catcode `\%12\relax}%
\providecommand \@@startlink[1]{}%
\providecommand \@@endlink[0]{}%
\providecommand \url  [0]{\begingroup\@sanitize@url \@url }%
\providecommand \@url [1]{\endgroup\@href {#1}{\urlprefix }}%
\providecommand \urlprefix  [0]{URL }%
\providecommand \Eprint [0]{\href }%
\providecommand \doibase [0]{https://doi.org/}%
\providecommand \selectlanguage [0]{\@gobble}%
\providecommand \bibinfo  [0]{\@secondoftwo}%
\providecommand \bibfield  [0]{\@secondoftwo}%
\providecommand \translation [1]{[#1]}%
\providecommand \BibitemOpen [0]{}%
\providecommand \bibitemStop [0]{}%
\providecommand \bibitemNoStop [0]{.\EOS\space}%
\providecommand \EOS [0]{\spacefactor3000\relax}%
\providecommand \BibitemShut  [1]{\csname bibitem#1\endcsname}%
\let\auto@bib@innerbib\@empty
\bibitem [{\citenamefont {Anderson}(1951)}]{Anderson}%
  \BibitemOpen
  \bibfield  {author} {\bibinfo {author} {\bibfnamefont {P.~W.}\ \bibnamefont
  {Anderson}},\ }\bibfield  {title} {\bibinfo {title} {Limits on the energy of
  the antiferromagnetic ground state},\ }\href
  {https://doi.org/10.1103/PhysRev.83.1260} {\bibfield  {journal} {\bibinfo
  {journal} {Phys. Rev.}\ }\textbf {\bibinfo {volume} {83}},\ \bibinfo {pages}
  {1260} (\bibinfo {year} {1951})}\BibitemShut {NoStop}%
\bibitem [{\citenamefont {Moessner}\ and\ \citenamefont
  {Chalker}(1998)}]{Moessner_PhysRevLett.80.2929}%
  \BibitemOpen
  \bibfield  {author} {\bibinfo {author} {\bibfnamefont {R.}~\bibnamefont
  {Moessner}}\ and\ \bibinfo {author} {\bibfnamefont {J.~T.}\ \bibnamefont
  {Chalker}},\ }\bibfield  {title} {\bibinfo {title} {Properties of a classical
  spin liquid: The heisenberg pyrochlore antiferromagnet},\ }\href
  {https://doi.org/10.1103/PhysRevLett.80.2929} {\bibfield  {journal} {\bibinfo
   {journal} {Phys. Rev. Lett.}\ }\textbf {\bibinfo {volume} {80}},\ \bibinfo
  {pages} {2929} (\bibinfo {year} {1998})}\BibitemShut {NoStop}%
\bibitem [{\citenamefont {Majumdar}\ and\ \citenamefont
  {Ghosh}(1969)}]{Majumdar69}%
  \BibitemOpen
  \bibfield  {author} {\bibinfo {author} {\bibfnamefont {C.~K.}\ \bibnamefont
  {Majumdar}}\ and\ \bibinfo {author} {\bibfnamefont {D.~K.}\ \bibnamefont
  {Ghosh}},\ }\bibfield  {title} {\bibinfo {title} {On
  next‐nearest‐neighbor interaction in linear chain. i},\ }\href
  {https://doi.org/10.1063/1.1664978} {\bibfield  {journal} {\bibinfo
  {journal} {Journal of Mathematical Physics}\ }\textbf {\bibinfo {volume}
  {10}},\ \bibinfo {pages} {1388} (\bibinfo {year} {1969})}\BibitemShut
  {NoStop}%
\bibitem [{\citenamefont {Majumdar}\ and\ \citenamefont
  {Shastry}(1976)}]{Majumdar76}%
  \BibitemOpen
  \bibfield  {author} {\bibinfo {author} {\bibfnamefont {C.~K.}\ \bibnamefont
  {Majumdar}}\ and\ \bibinfo {author} {\bibfnamefont {B.~S.}\ \bibnamefont
  {Shastry}},\ }\bibfield  {title} {\bibinfo {title} {Lower bound to the
  ground-state energy of the heisenberg hamiltonian on a triangular lattice},\
  }\href {https://doi.org/10.1080/14786437608221128} {\bibfield  {journal}
  {\bibinfo  {journal} {The Philosophical Magazine: A Journal of Theoretical
  Experimental and Applied Physics}\ }\textbf {\bibinfo {volume} {33}},\
  \bibinfo {pages} {685} (\bibinfo {year} {1976})},\ \Eprint
  {https://arxiv.org/abs/https://doi.org/10.1080/14786437608221128}
  {https://doi.org/10.1080/14786437608221128} \BibitemShut {NoStop}%
\bibitem [{\citenamefont {Tarrach}\ and\ \citenamefont
  {Valent\'{\i}}(1990)}]{Tarrach90}%
  \BibitemOpen
  \bibfield  {author} {\bibinfo {author} {\bibfnamefont {R.}~\bibnamefont
  {Tarrach}}\ and\ \bibinfo {author} {\bibfnamefont {R.}~\bibnamefont
  {Valent\'{\i}}},\ }\bibfield  {title} {\bibinfo {title} {Exact lower bounds
  to the ground-state energy of spin systems: The two-dimensional s=1/2
  antiferromagnetic heisenberg model},\ }\href
  {https://doi.org/10.1103/PhysRevB.41.9611} {\bibfield  {journal} {\bibinfo
  {journal} {Phys. Rev. B}\ }\textbf {\bibinfo {volume} {41}},\ \bibinfo
  {pages} {9611} (\bibinfo {year} {1990})}\BibitemShut {NoStop}%
\bibitem [{\citenamefont {Nishimori}\ and\ \citenamefont
  {Ozeki}(1989)}]{Nishimori89}%
  \BibitemOpen
  \bibfield  {author} {\bibinfo {author} {\bibfnamefont {H.}~\bibnamefont
  {Nishimori}}\ and\ \bibinfo {author} {\bibfnamefont {Y.}~\bibnamefont
  {Ozeki}},\ }\bibfield  {title} {\bibinfo {title} {Ground-state long-range
  order in the two-dimensional xxz mode},\ }\href
  {https://doi.org/10.1143/JPSJ.58.1027} {\bibfield  {journal} {\bibinfo
  {journal} {Journal of the Physical Society of Japan}\ }\textbf {\bibinfo
  {volume} {58}},\ \bibinfo {pages} {1027} (\bibinfo {year} {1989})},\ \Eprint
  {https://arxiv.org/abs/https://doi.org/10.1143/JPSJ.58.1027}
  {https://doi.org/10.1143/JPSJ.58.1027} \BibitemShut {NoStop}%
\bibitem [{\citenamefont {Barthel}\ and\ \citenamefont
  {H\"ubener}(2012)}]{Barthel2012}%
  \BibitemOpen
  \bibfield  {author} {\bibinfo {author} {\bibfnamefont {T.}~\bibnamefont
  {Barthel}}\ and\ \bibinfo {author} {\bibfnamefont {R.}~\bibnamefont
  {H\"ubener}},\ }\bibfield  {title} {\bibinfo {title} {Solving
  condensed-matter ground-state problems by semidefinite relaxations},\ }\href
  {https://doi.org/10.1103/PhysRevLett.108.200404} {\bibfield  {journal}
  {\bibinfo  {journal} {Phys. Rev. Lett.}\ }\textbf {\bibinfo {volume} {108}},\
  \bibinfo {pages} {200404} (\bibinfo {year} {2012})}\BibitemShut {NoStop}%
\bibitem [{\citenamefont {Han}(2020)}]{han2020}%
  \BibitemOpen
  \bibfield  {author} {\bibinfo {author} {\bibfnamefont {X.}~\bibnamefont
  {Han}},\ }\href {https://arxiv.org/abs/2006.06002} {\bibinfo {title} {Quantum
  many-body bootstrap}} (\bibinfo {year} {2020}),\ \Eprint
  {https://arxiv.org/abs/2006.06002} {arXiv:2006.06002 [cond-mat.str-el]}
  \BibitemShut {NoStop}%
\bibitem [{\citenamefont {Requena}\ \emph {et~al.}(2023)\citenamefont
  {Requena}, \citenamefont {Mu\~noz Gil}, \citenamefont {Lewenstein},
  \citenamefont {Dunjko},\ and\ \citenamefont {Tura}}]{Requena2023}%
  \BibitemOpen
  \bibfield  {author} {\bibinfo {author} {\bibfnamefont {B.}~\bibnamefont
  {Requena}}, \bibinfo {author} {\bibfnamefont {G.}~\bibnamefont {Mu\~noz
  Gil}}, \bibinfo {author} {\bibfnamefont {M.}~\bibnamefont {Lewenstein}},
  \bibinfo {author} {\bibfnamefont {V.}~\bibnamefont {Dunjko}},\ and\ \bibinfo
  {author} {\bibfnamefont {J.}~\bibnamefont {Tura}},\ }\bibfield  {title}
  {\bibinfo {title} {Certificates of quantum many-body properties assisted by
  machine learning},\ }\href {https://doi.org/10.1103/PhysRevResearch.5.013097}
  {\bibfield  {journal} {\bibinfo  {journal} {Phys. Rev. Res.}\ }\textbf
  {\bibinfo {volume} {5}},\ \bibinfo {pages} {013097} (\bibinfo {year}
  {2023})}\BibitemShut {NoStop}%
\bibitem [{\citenamefont {Wang}\ \emph {et~al.}(2024)\citenamefont {Wang},
  \citenamefont {Surace}, \citenamefont {Fr\'erot}, \citenamefont {Legat},
  \citenamefont {Renou}, \citenamefont {Magron},\ and\ \citenamefont
  {Ac\'{\i}n}}]{Wang2024}%
  \BibitemOpen
  \bibfield  {author} {\bibinfo {author} {\bibfnamefont {J.}~\bibnamefont
  {Wang}}, \bibinfo {author} {\bibfnamefont {J.}~\bibnamefont {Surace}},
  \bibinfo {author} {\bibfnamefont {I.}~\bibnamefont {Fr\'erot}}, \bibinfo
  {author} {\bibfnamefont {B.}~\bibnamefont {Legat}}, \bibinfo {author}
  {\bibfnamefont {M.-O.}\ \bibnamefont {Renou}}, \bibinfo {author}
  {\bibfnamefont {V.}~\bibnamefont {Magron}},\ and\ \bibinfo {author}
  {\bibfnamefont {A.}~\bibnamefont {Ac\'{\i}n}},\ }\bibfield  {title} {\bibinfo
  {title} {Certifying ground-state properties of many-body systems},\ }\href
  {https://doi.org/10.1103/PhysRevX.14.031006} {\bibfield  {journal} {\bibinfo
  {journal} {Phys. Rev. X}\ }\textbf {\bibinfo {volume} {14}},\ \bibinfo
  {pages} {031006} (\bibinfo {year} {2024})}\BibitemShut {NoStop}%
\bibitem [{\citenamefont {Kull}\ \emph {et~al.}(2024)\citenamefont {Kull},
  \citenamefont {Schuch}, \citenamefont {Dive},\ and\ \citenamefont
  {Navascu\'es}}]{Kull2024}%
  \BibitemOpen
  \bibfield  {author} {\bibinfo {author} {\bibfnamefont {I.}~\bibnamefont
  {Kull}}, \bibinfo {author} {\bibfnamefont {N.}~\bibnamefont {Schuch}},
  \bibinfo {author} {\bibfnamefont {B.}~\bibnamefont {Dive}},\ and\ \bibinfo
  {author} {\bibfnamefont {M.}~\bibnamefont {Navascu\'es}},\ }\bibfield
  {title} {\bibinfo {title} {Lower bounds on ground-state energies of local
  hamiltonians through the renormalization group},\ }\href
  {https://doi.org/10.1103/PhysRevX.14.021008} {\bibfield  {journal} {\bibinfo
  {journal} {Phys. Rev. X}\ }\textbf {\bibinfo {volume} {14}},\ \bibinfo
  {pages} {021008} (\bibinfo {year} {2024})}\BibitemShut {NoStop}%
\bibitem [{\citenamefont {Canals}\ and\ \citenamefont
  {Lacroix}(2000)}]{canals}%
  \BibitemOpen
  \bibfield  {author} {\bibinfo {author} {\bibfnamefont {B.}~\bibnamefont
  {Canals}}\ and\ \bibinfo {author} {\bibfnamefont {C.}~\bibnamefont
  {Lacroix}},\ }\bibfield  {title} {\bibinfo {title} {Quantum spin liquid: The
  heisenberg antiferromagnet on the three-dimensional pyrochlore lattice},\
  }\href {https://doi.org/10.1103/PhysRevB.61.1149} {\bibfield  {journal}
  {\bibinfo  {journal} {Phys. Rev. B}\ }\textbf {\bibinfo {volume} {61}},\
  \bibinfo {pages} {1149} (\bibinfo {year} {2000})}\BibitemShut {NoStop}%
\bibitem [{\citenamefont {Iqbal}\ \emph {et~al.}(2019)\citenamefont {Iqbal},
  \citenamefont {M\"uller}, \citenamefont {Ghosh}, \citenamefont {Gingras},
  \citenamefont {Jeschke}, \citenamefont {Rachel}, \citenamefont {Reuther},\
  and\ \citenamefont {Thomale}}]{yasir}%
  \BibitemOpen
  \bibfield  {author} {\bibinfo {author} {\bibfnamefont {Y.}~\bibnamefont
  {Iqbal}}, \bibinfo {author} {\bibfnamefont {T.}~\bibnamefont {M\"uller}},
  \bibinfo {author} {\bibfnamefont {P.}~\bibnamefont {Ghosh}}, \bibinfo
  {author} {\bibfnamefont {M.~J.~P.}\ \bibnamefont {Gingras}}, \bibinfo
  {author} {\bibfnamefont {H.~O.}\ \bibnamefont {Jeschke}}, \bibinfo {author}
  {\bibfnamefont {S.}~\bibnamefont {Rachel}}, \bibinfo {author} {\bibfnamefont
  {J.}~\bibnamefont {Reuther}},\ and\ \bibinfo {author} {\bibfnamefont
  {R.}~\bibnamefont {Thomale}},\ }\bibfield  {title} {\bibinfo {title} {Quantum
  and classical phases of the pyrochlore heisenberg model with competing
  interactions},\ }\href {https://doi.org/10.1103/PhysRevX.9.011005} {\bibfield
   {journal} {\bibinfo  {journal} {Phys. Rev. X}\ }\textbf {\bibinfo {volume}
  {9}},\ \bibinfo {pages} {011005} (\bibinfo {year} {2019})}\BibitemShut
  {NoStop}%
\bibitem [{\citenamefont {Pohle}\ \emph {et~al.}(2023)\citenamefont {Pohle},
  \citenamefont {Yamaji},\ and\ \citenamefont {Imada}}]{pohle2023}%
  \BibitemOpen
  \bibfield  {author} {\bibinfo {author} {\bibfnamefont {R.}~\bibnamefont
  {Pohle}}, \bibinfo {author} {\bibfnamefont {Y.}~\bibnamefont {Yamaji}},\ and\
  \bibinfo {author} {\bibfnamefont {M.}~\bibnamefont {Imada}},\ }\href
  {https://arxiv.org/abs/2311.11561} {\bibinfo {title} {Ground state of the
  $s$=1/2 pyrochlore heisenberg antiferromagnet: A quantum spin liquid emergent
  from dimensional reduction}} (\bibinfo {year} {2023}),\ \Eprint
  {https://arxiv.org/abs/2311.11561} {arXiv:2311.11561 [cond-mat.str-el]}
  \BibitemShut {NoStop}%
\bibitem [{\citenamefont {Hering}\ \emph {et~al.}(2022)\citenamefont {Hering},
  \citenamefont {Noculak}, \citenamefont {Ferrari}, \citenamefont {Iqbal},\
  and\ \citenamefont {Reuther}}]{hering2022}%
  \BibitemOpen
  \bibfield  {author} {\bibinfo {author} {\bibfnamefont {M.}~\bibnamefont
  {Hering}}, \bibinfo {author} {\bibfnamefont {V.}~\bibnamefont {Noculak}},
  \bibinfo {author} {\bibfnamefont {F.}~\bibnamefont {Ferrari}}, \bibinfo
  {author} {\bibfnamefont {Y.}~\bibnamefont {Iqbal}},\ and\ \bibinfo {author}
  {\bibfnamefont {J.}~\bibnamefont {Reuther}},\ }\bibfield  {title} {\bibinfo
  {title} {Dimerization tendencies of the pyrochlore heisenberg
  antiferromagnet: A functional renormalization group perspective},\ }\href
  {https://doi.org/10.1103/PhysRevB.105.054426} {\bibfield  {journal} {\bibinfo
   {journal} {Phys. Rev. B}\ }\textbf {\bibinfo {volume} {105}},\ \bibinfo
  {pages} {054426} (\bibinfo {year} {2022})}\BibitemShut {NoStop}%
\bibitem [{\citenamefont {Hagym\'asi}\ \emph {et~al.}(2021)\citenamefont
  {Hagym\'asi}, \citenamefont {Sch\"afer}, \citenamefont {Moessner},\ and\
  \citenamefont {Luitz}}]{hagymasi}%
  \BibitemOpen
  \bibfield  {author} {\bibinfo {author} {\bibfnamefont {I.}~\bibnamefont
  {Hagym\'asi}}, \bibinfo {author} {\bibfnamefont {R.}~\bibnamefont
  {Sch\"afer}}, \bibinfo {author} {\bibfnamefont {R.}~\bibnamefont
  {Moessner}},\ and\ \bibinfo {author} {\bibfnamefont {D.~J.}\ \bibnamefont
  {Luitz}},\ }\bibfield  {title} {\bibinfo {title} {Possible inversion symmetry
  breaking in the $s=1/2$ pyrochlore heisenberg magnet},\ }\href
  {https://doi.org/10.1103/PhysRevLett.126.117204} {\bibfield  {journal}
  {\bibinfo  {journal} {Phys. Rev. Lett.}\ }\textbf {\bibinfo {volume} {126}},\
  \bibinfo {pages} {117204} (\bibinfo {year} {2021})}\BibitemShut {NoStop}%
\bibitem [{\citenamefont {Astrakhantsev}\ \emph {et~al.}(2021)\citenamefont
  {Astrakhantsev}, \citenamefont {Westerhout}, \citenamefont {Tiwari},
  \citenamefont {Choo}, \citenamefont {Chen}, \citenamefont {Fischer},
  \citenamefont {Carleo},\ and\ \citenamefont {Neupert}}]{astrakhantsev2021}%
  \BibitemOpen
  \bibfield  {author} {\bibinfo {author} {\bibfnamefont {N.}~\bibnamefont
  {Astrakhantsev}}, \bibinfo {author} {\bibfnamefont {T.}~\bibnamefont
  {Westerhout}}, \bibinfo {author} {\bibfnamefont {A.}~\bibnamefont {Tiwari}},
  \bibinfo {author} {\bibfnamefont {K.}~\bibnamefont {Choo}}, \bibinfo {author}
  {\bibfnamefont {A.}~\bibnamefont {Chen}}, \bibinfo {author} {\bibfnamefont
  {M.~H.}\ \bibnamefont {Fischer}}, \bibinfo {author} {\bibfnamefont
  {G.}~\bibnamefont {Carleo}},\ and\ \bibinfo {author} {\bibfnamefont
  {T.}~\bibnamefont {Neupert}},\ }\bibfield  {title} {\bibinfo {title}
  {Broken-symmetry ground states of the heisenberg model on the pyrochlore
  lattice},\ }\href {https://doi.org/10.1103/PhysRevX.11.041021} {\bibfield
  {journal} {\bibinfo  {journal} {Phys. Rev. X}\ }\textbf {\bibinfo {volume}
  {11}},\ \bibinfo {pages} {041021} (\bibinfo {year} {2021})}\BibitemShut
  {NoStop}%
\bibitem [{\citenamefont {Sch\"afer}\ \emph {et~al.}(2023)\citenamefont
  {Sch\"afer}, \citenamefont {Placke}, \citenamefont {Benton},\ and\
  \citenamefont {Moessner}}]{schafer}%
  \BibitemOpen
  \bibfield  {author} {\bibinfo {author} {\bibfnamefont {R.}~\bibnamefont
  {Sch\"afer}}, \bibinfo {author} {\bibfnamefont {B.}~\bibnamefont {Placke}},
  \bibinfo {author} {\bibfnamefont {O.}~\bibnamefont {Benton}},\ and\ \bibinfo
  {author} {\bibfnamefont {R.}~\bibnamefont {Moessner}},\ }\bibfield  {title}
  {\bibinfo {title} {Abundance of hard-hexagon crystals in the quantum
  pyrochlore antiferromagnet},\ }\href
  {https://doi.org/10.1103/PhysRevLett.131.096702} {\bibfield  {journal}
  {\bibinfo  {journal} {Phys. Rev. Lett.}\ }\textbf {\bibinfo {volume} {131}},\
  \bibinfo {pages} {096702} (\bibinfo {year} {2023})}\BibitemShut {NoStop}%
\bibitem [{\citenamefont {Cheng}\ and\ \citenamefont {Li}(2025)}]{cheng2025}%
  \BibitemOpen
  \bibfield  {author} {\bibinfo {author} {\bibfnamefont {R.}~\bibnamefont
  {Cheng}}\ and\ \bibinfo {author} {\bibfnamefont {T.}~\bibnamefont {Li}},\
  }\href {https://arxiv.org/abs/2509.13746} {\bibinfo {title} {Closely
  competing valence bond crystal orders in the ground state of the
  spin-$\frac{1}{2}$ antiferromagnetic heisenberg model on the pyrochlore
  lattice: a large scale unrestricted variational study}} (\bibinfo {year}
  {2025}),\ \Eprint {https://arxiv.org/abs/2509.13746} {arXiv:2509.13746
  [cond-mat.str-el]} \BibitemShut {NoStop}%
\bibitem [{\citenamefont {Richter}\ \emph {et~al.}(2004)\citenamefont
  {Richter}, \citenamefont {Schulenburg},\ and\ \citenamefont
  {Honecker}}]{richter2004}%
  \BibitemOpen
  \bibfield  {author} {\bibinfo {author} {\bibfnamefont {J.}~\bibnamefont
  {Richter}}, \bibinfo {author} {\bibfnamefont {J.}~\bibnamefont
  {Schulenburg}},\ and\ \bibinfo {author} {\bibfnamefont {A.}~\bibnamefont
  {Honecker}},\ }\bibfield  {title} {\bibinfo {title} {Quantum magnetism in two
  dimensions: From semi-classical n\'eel order to magnetic disorder},\ }in\
  \href@noop {} {\emph {\bibinfo {booktitle} {Quantum Magnetism}}},\ \bibinfo
  {series} {Lecture Notes in Physics}, Vol.\ \bibinfo {volume} {645}\ (\bibinfo
   {publisher} {Springer},\ \bibinfo {year} {2004})\ pp.\ \bibinfo {pages}
  {85--153}\BibitemShut {NoStop}%
\bibitem [{\citenamefont {{Richter}}\ \emph {et~al.}(2004)\citenamefont
  {{Richter}}, \citenamefont {{Schulenburg}}, \citenamefont {{Honecker}},
  \citenamefont {{Schnack}},\ and\ \citenamefont
  {{Schmidt}}}]{2004JPCM...16S.779R}%
  \BibitemOpen
  \bibfield  {author} {\bibinfo {author} {\bibfnamefont {J.}~\bibnamefont
  {{Richter}}}, \bibinfo {author} {\bibfnamefont {J.}~\bibnamefont
  {{Schulenburg}}}, \bibinfo {author} {\bibfnamefont {A.}~\bibnamefont
  {{Honecker}}}, \bibinfo {author} {\bibfnamefont {J.}~\bibnamefont
  {{Schnack}}},\ and\ \bibinfo {author} {\bibfnamefont {H.-J.}\ \bibnamefont
  {{Schmidt}}},\ }\bibfield  {title} {\bibinfo {title} {{Exact eigenstates and
  macroscopic magnetization jumps in strongly frustrated spin lattices}},\
  }\href {https://doi.org/10.1088/0953-8984/16/11/029} {\bibfield  {journal}
  {\bibinfo  {journal} {Journal of Physics Condensed Matter}\ }\textbf
  {\bibinfo {volume} {16}},\ \bibinfo {pages} {S779} (\bibinfo {year}
  {2004})},\ \Eprint {https://arxiv.org/abs/cond-mat/0309455}
  {arXiv:cond-mat/0309455 [cond-mat.str-el]} \BibitemShut {NoStop}%
\bibitem [{\citenamefont {Hahn}(2016)}]{ITC_A}%
  \BibitemOpen
  \bibinfo {editor} {\bibfnamefont {T.}~\bibnamefont {Hahn}},\ ed.,\ \href@noop
  {} {\emph {\bibinfo {title} {International Tables for Crystallography, Volume
  A: Space-Group Symmetry}}},\ \bibinfo {edition} {6th}\ ed.\ (\bibinfo
  {publisher} {International Union of Crystallography},\ \bibinfo {address}
  {Chester, England},\ \bibinfo {year} {2016})\BibitemShut {NoStop}%
\bibitem [{\citenamefont {Lieb}\ and\ \citenamefont {Mattis}(1962)}]{Lieb}%
  \BibitemOpen
  \bibfield  {author} {\bibinfo {author} {\bibfnamefont {E.}~\bibnamefont
  {Lieb}}\ and\ \bibinfo {author} {\bibfnamefont {D.}~\bibnamefont {Mattis}},\
  }\bibfield  {title} {\bibinfo {title} {{Ordering Energy Levels of Interacting
  Spin Systems}},\ }\href {https://doi.org/10.1063/1.1724276} {\bibfield
  {journal} {\bibinfo  {journal} {Journal of Mathematical Physics}\ }\textbf
  {\bibinfo {volume} {3}},\ \bibinfo {pages} {749} (\bibinfo {year} {1962})},\
  \Eprint
  {https://arxiv.org/abs/https://pubs.aip.org/aip/jmp/article-pdf/3/4/749/19167430/749\_1\_online.pdf}
  {https://pubs.aip.org/aip/jmp/article-pdf/3/4/749/19167430/749\_1\_online.pdf}
  \BibitemShut {NoStop}%
\bibitem [{\citenamefont {Nussinov}\ \emph {et~al.}(2007)\citenamefont
  {Nussinov}, \citenamefont {Batista}, \citenamefont {Normand},\ and\
  \citenamefont {Trugman}}]{nussinov2007}%
  \BibitemOpen
  \bibfield  {author} {\bibinfo {author} {\bibfnamefont {Z.}~\bibnamefont
  {Nussinov}}, \bibinfo {author} {\bibfnamefont {C.~D.}\ \bibnamefont
  {Batista}}, \bibinfo {author} {\bibfnamefont {B.}~\bibnamefont {Normand}},\
  and\ \bibinfo {author} {\bibfnamefont {S.~A.}\ \bibnamefont {Trugman}},\
  }\bibfield  {title} {\bibinfo {title} {High-dimensional fractionalization and
  spinon deconfinement in pyrochlore antiferromagnets},\ }\href
  {https://doi.org/10.1103/PhysRevB.75.094411} {\bibfield  {journal} {\bibinfo
  {journal} {Phys. Rev. B}\ }\textbf {\bibinfo {volume} {75}},\ \bibinfo
  {pages} {094411} (\bibinfo {year} {2007})}\BibitemShut {NoStop}%
\bibitem [{\citenamefont {Lozano-G{\'o}mez}\ \emph {et~al.}(2024)\citenamefont
  {Lozano-G{\'o}mez}, \citenamefont {Iqbal},\ and\ \citenamefont
  {Vojta}}]{Lozano2024}%
  \BibitemOpen
  \bibfield  {author} {\bibinfo {author} {\bibfnamefont {D.}~\bibnamefont
  {Lozano-G{\'o}mez}}, \bibinfo {author} {\bibfnamefont {Y.}~\bibnamefont
  {Iqbal}},\ and\ \bibinfo {author} {\bibfnamefont {M.}~\bibnamefont {Vojta}},\
  }\bibfield  {title} {\bibinfo {title} {A classical chiral spin liquid from
  chiral interactions on the pyrochlore lattice},\ }\href
  {https://doi.org/10.1038/s41467-024-54558-7} {\bibfield  {journal} {\bibinfo
  {journal} {Nat. Commun.}\ }\textbf {\bibinfo {volume} {15}},\ \bibinfo
  {pages} {10162} (\bibinfo {year} {2024})}\BibitemShut {NoStop}%
\bibitem [{\citenamefont {Pitts}\ \emph {et~al.}(2022)\citenamefont {Pitts},
  \citenamefont {Buessen}, \citenamefont {Moessner}, \citenamefont {Trebst},\
  and\ \citenamefont {Shtengel}}]{Pitts2022}%
  \BibitemOpen
  \bibfield  {author} {\bibinfo {author} {\bibfnamefont {J.}~\bibnamefont
  {Pitts}}, \bibinfo {author} {\bibfnamefont {F.~L.}\ \bibnamefont {Buessen}},
  \bibinfo {author} {\bibfnamefont {R.}~\bibnamefont {Moessner}}, \bibinfo
  {author} {\bibfnamefont {S.}~\bibnamefont {Trebst}},\ and\ \bibinfo {author}
  {\bibfnamefont {K.}~\bibnamefont {Shtengel}},\ }\bibfield  {title} {\bibinfo
  {title} {Order by disorder in classical kagome antiferromagnets with chiral
  interactions},\ }\href {https://doi.org/10.1103/PhysRevResearch.4.043019}
  {\bibfield  {journal} {\bibinfo  {journal} {Phys. Rev. Res.}\ }\textbf
  {\bibinfo {volume} {4}},\ \bibinfo {pages} {043019} (\bibinfo {year}
  {2022})}\BibitemShut {NoStop}%
\bibitem [{\citenamefont {Oliviero}\ \emph {et~al.}(2022)\citenamefont
  {Oliviero}, \citenamefont {Sobral}, \citenamefont {Andrade},\ and\
  \citenamefont {Pereira}}]{Fabrizio2022}%
  \BibitemOpen
  \bibfield  {author} {\bibinfo {author} {\bibfnamefont {F.}~\bibnamefont
  {Oliviero}}, \bibinfo {author} {\bibfnamefont {J.~A.}\ \bibnamefont
  {Sobral}}, \bibinfo {author} {\bibfnamefont {E.~C.}\ \bibnamefont
  {Andrade}},\ and\ \bibinfo {author} {\bibfnamefont {R.~G.}\ \bibnamefont
  {Pereira}},\ }\bibfield  {title} {\bibinfo {title} {{Noncoplanar magnetic
  orders and gapless chiral spin liquid on the kagome lattice with staggered
  scalar spin chirality}},\ }\href
  {https://doi.org/10.21468/SciPostPhys.13.3.050} {\bibfield  {journal}
  {\bibinfo  {journal} {SciPost Phys.}\ }\textbf {\bibinfo {volume} {13}},\
  \bibinfo {pages} {050} (\bibinfo {year} {2022})}\BibitemShut {NoStop}%
\bibitem [{\citenamefont {Xian}(1995)}]{Xian_PRB.52.12485_couple_chains}%
  \BibitemOpen
  \bibfield  {author} {\bibinfo {author} {\bibfnamefont {Y.}~\bibnamefont
  {Xian}},\ }\bibfield  {title} {\bibinfo {title} {Rigorous results on a
  first-order phase transition in antiferromagnetic spin-1/2 coupled chains},\
  }\href {https://doi.org/10.1103/PhysRevB.52.12485} {\bibfield  {journal}
  {\bibinfo  {journal} {Phys. Rev. B}\ }\textbf {\bibinfo {volume} {52}},\
  \bibinfo {pages} {12485} (\bibinfo {year} {1995})}\BibitemShut {NoStop}%
\bibitem [{\citenamefont {Honecker}\ \emph {et~al.}(2000)\citenamefont
  {Honecker}, \citenamefont {Mila},\ and\ \citenamefont
  {Troyer}}]{Honecker_Mila_Troyer_2000}%
  \BibitemOpen
  \bibfield  {author} {\bibinfo {author} {\bibfnamefont {A.}~\bibnamefont
  {Honecker}}, \bibinfo {author} {\bibfnamefont {F.}~\bibnamefont {Mila}},\
  and\ \bibinfo {author} {\bibfnamefont {M.}~\bibnamefont {Troyer}},\
  }\bibfield  {title} {\bibinfo {title} {Magnetization plateaux and jumps in a
  class of frustrated ladders: A simple route to a complex behaviour},\ }\href
  {https://doi.org/10.1007/s100510051120} {\bibfield  {journal} {\bibinfo
  {journal} {The European Physical Journal B - Condensed Matter and Complex
  Systems}\ }\textbf {\bibinfo {volume} {15}},\ \bibinfo {pages} {227}
  (\bibinfo {year} {2000})}\BibitemShut {NoStop}%
\bibitem [{\citenamefont {Lamas}\ and\ \citenamefont
  {Matera}(2015)}]{Lamas_Matera_PRB_92_Dimerized}%
  \BibitemOpen
  \bibfield  {author} {\bibinfo {author} {\bibfnamefont {C.~A.}\ \bibnamefont
  {Lamas}}\ and\ \bibinfo {author} {\bibfnamefont {J.~M.}\ \bibnamefont
  {Matera}},\ }\bibfield  {title} {\bibinfo {title} {Dimerized ground states in
  $\text{spin}\ensuremath{-}s$ frustrated systems},\ }\href
  {https://doi.org/10.1103/PhysRevB.92.115111} {\bibfield  {journal} {\bibinfo
  {journal} {Phys. Rev. B}\ }\textbf {\bibinfo {volume} {92}},\ \bibinfo
  {pages} {115111} (\bibinfo {year} {2015})}\BibitemShut {NoStop}%
\bibitem [{\citenamefont {Sobral}\ and\ \citenamefont
  {Lacroix}(1997)}]{sobral}%
  \BibitemOpen
  \bibfield  {author} {\bibinfo {author} {\bibfnamefont {R.}~\bibnamefont
  {Sobral}}\ and\ \bibinfo {author} {\bibfnamefont {C.}~\bibnamefont
  {Lacroix}},\ }\bibfield  {title} {\bibinfo {title} {Order by disorder in the
  pyrochlore antiferromagnets},\ }\href
  {https://doi.org/https://doi.org/10.1016/S0038-1098(97)00212-3} {\bibfield
  {journal} {\bibinfo  {journal} {Solid State Communications}\ }\textbf
  {\bibinfo {volume} {103}},\ \bibinfo {pages} {407} (\bibinfo {year}
  {1997})}\BibitemShut {NoStop}%
\bibitem [{\citenamefont {Derzhko}\ \emph {et~al.}(2020)\citenamefont
  {Derzhko}, \citenamefont {Hutak}, \citenamefont {Krokhmalskii}, \citenamefont
  {Schnack},\ and\ \citenamefont {Richter}}]{drezhko}%
  \BibitemOpen
  \bibfield  {author} {\bibinfo {author} {\bibfnamefont {O.}~\bibnamefont
  {Derzhko}}, \bibinfo {author} {\bibfnamefont {T.}~\bibnamefont {Hutak}},
  \bibinfo {author} {\bibfnamefont {T.}~\bibnamefont {Krokhmalskii}}, \bibinfo
  {author} {\bibfnamefont {J.}~\bibnamefont {Schnack}},\ and\ \bibinfo {author}
  {\bibfnamefont {J.}~\bibnamefont {Richter}},\ }\bibfield  {title} {\bibinfo
  {title} {Adapting planck's route to investigate the thermodynamics of the
  spin-half pyrochlore heisenberg antiferromagnet},\ }\href
  {https://doi.org/10.1103/PhysRevB.101.174426} {\bibfield  {journal} {\bibinfo
   {journal} {Phys. Rev. B}\ }\textbf {\bibinfo {volume} {101}},\ \bibinfo
  {pages} {174426} (\bibinfo {year} {2020})}\BibitemShut {NoStop}%
\end{thebibliography}%

\end{document}